%% file: main.tex
\documentclass[twocolumn]{aastex63}
\usepackage{amsmath}
\usepackage{listings}
\usepackage{xcolor}

\definecolor{codegreen}{rgb}{0,0.6,0}
\definecolor{codegray}{rgb}{0.5,0.5,0.5}
\definecolor{codepurple}{rgb}{0.58,0,0.82}
\definecolor{backcolour}{rgb}{0.95,0.95,0.92}

\defcitealias{Lohner-Bottcher2019}{L19}
\defcitealias{Lohner-Bottcher2018}{L18}

\newcommand{\ms}{{\rm m\ s}^{-1}}
\newcommand{\cms}{\ {\rm cm\ s}^{-1}}

\newcommand{\Halpha}{H$\alpha$}
\newcommand{\grass}{\texttt{GRASS}}
\newcommand{\PSUAA}{Department of Astronomy \& Astrophysics, 525 Davey Laboratory, The Pennsylvania State University, University Park, PA, 16802, USA}
\newcommand{\PSUCEHW}{Center for Exoplanets and Habitable Worlds, 525 Davey Laboratory, The Pennsylvania State University, University Park, PA, 16802, USA}
\newcommand{\PSETI}{Penn State Extraterrestrial Intelligence Center, 525 Davey Laboratory, The Pennsylvania State University, University Park, PA, 16802, USA}
\newcommand{\PSUStats}{Center for Astrostatistics, The Pennsylvania State University, University Park, PA, 16802, USA}
\newcommand{\PSUCDS}{Institute for Computational and Data Sciences, The Pennsylvania State University, 525 Davey Laboratory, University Park, PA 16802, USA}

\definecolor{ebf}{rgb}{0.4, 0.0, 0.6}

\definecolor{mlp}{rgb}{0.294, 0.612, 0.827}

\newcommand{\revise}[1]{#1}

\newcommand{\revisetwo}[1]{#1}

\usepackage{natbib}
\shorttitle{GRASS: Granulation and Spectrum Simulator}
\shortauthors{M.L. Palumbo III et al.}

\graphicspath{{./}{figures/}}
\begin{document}
\tabletypesize{\small}

\title{GRASS: Distinguishing Planet-induced Doppler Signatures from Granulation with a Synthetic Spectra Generator}

\author[0000-0002-4677-8796]{Michael L. Palumbo III}
\affiliation{\PSUAA}
\affiliation{\PSUCEHW}

\author[0000-0001-6545-639X]{Eric B. Ford}
\affiliation{\PSUAA}
\affiliation{\PSUCEHW}
\affiliation{\PSUCDS}
\affiliation{\PSUStats}
\affiliation{Institute for Advanced Study, 1 Einstein Drive, Princeton, NJ, 08540, USA}

\author[0000-0001-6160-5888]{Jason T. Wright}
\affiliation{\PSUAA}
\affiliation{\PSUCEHW}
\affiliation{\PSETI}

\author[0000-0001-9596-7983]{Suvrath Mahadevan}
\affiliation{\PSUAA}
\affiliation{\PSUCEHW}

\author[0000-0002-5013-5769]{Alexander W. Wise}
\affiliation{\PSUAA}
\affiliation{\PSUCEHW}

\author{Johannes L{\"o}hner-B{\"o}ttcher}
\affiliation{High-Altitude Observatory, National Center for Atmospheric Research, 3080 Center Green Drive, Boulder, CO 80301, USA}

\correspondingauthor{Michael L. Palumbo III}
\email{palumbo@psu.edu}

\begin{abstract}
Owing to recent advances in radial-velocity instrumentation and observation techniques, the detection of Earth-mass planets around Sun-like stars may soon be primarily limited by intrinsic stellar variability. Several processes contribute to this variability, including starspots, pulsations, and granulation. Although many previous studies have focused on techniques to mitigate signals from pulsations and other types of magnetic activity, granulation noise has to date only been partially addressed by empirically-motivated observation strategies and magnetohydrodynamic simulations. To address this deficit, we present the GRanulation And Spectrum Simulator (\grass), a new tool designed to create time-series synthetic spectra \revise{with granulation-driven variability} from spatially- and temporally-resolved observations of solar \revise{absorption lines.} In this work, we present \grass, detail its methodology, and validate its model against disk-integrated solar observations. As a first-of-its-kind empirical model for spectral variability due to granulation in a star with perfectly known center-of-mass radial-velocity behavior, \grass\ is an important tool for testing new methods of disentangling granular line-shape changes from true Doppler shifts. 
\end{abstract}


\section{Introduction} \label{intro}
Since the detection of the first exoplanet orbiting a main sequence star \citep{Mayor1995}, the radial velocity (RV) method has enabled the discovery of several hundred planets to date \citep[see the NASA Exoplanet Archive;][]{Akeson2017}. Despite these successes, the RV method will \revise{need additional development to address stellar variability} and characterize Earth-analog planets around solar-type stars. In principle, the RV method uses the planet-induced Doppler reflex motion of host stars to infer the existence of planetary companions. In practice, the host stars exhibit intrinsic astrophysical variability that complicate or preclude the measurement of minute reflex velocities due to smaller, Earth-mass planets. Although advances in instrumentation have dramatically improved spectrograph stability and spectral sensitivity, the challenges posed by intrinsic stellar variability remain a significant barrier to the discovery of lower-mass, rocky planets \revise{in or near the habitable zone of other stellar systems.} \par

Several physical mechanisms, which operate on many \revise{different} timescales, drive stellar variability, including pulsations, magnetic activity, and granulation \revise{(for a detailed review of these and other phenomena see \citealt{Meunier2021})}. Pulsations refer to the rich spectrum of acoustic modes excited by instabilities in near-surface convection. For dwarf F-, G-, and K-type (FGK) stars, the principal pulsations occur on timescales of minutes and lead to residual Doppler velocity amplitudes large enough to be problematic for the search for Earth-like exoplanets (i.e., \revise{multiple} tens of centimeters per second \revise{from pulsations alone, compared to the $\sim$10$\cms$ signal expected for an Earth-twin}; e.g., \citealt{O'Toole2008} and \citealt{Michel2008}). Recently, \citet{Chaplin2019} have shown that carefully-planned exposure times equal to or greater than the period of the principal $p$-mode can effectively mitigate the observed pulsation amplitude down to the $0.1\ \ms$ level necessary for Earth-twin detection. \par

In comparison to pulsations alone, stellar magnetic activity has posed a relatively complex barrier. Starspots, cooler photospheric regions where penetrating magnetic fields inhibit surface convection, produce a rotationally-modulated, wavelength-dependent flux deficit that alters the shape of spectral lines and can therefore mimic the Doppler wobble induced by a planetary companion. In the chromosphere, concentrated magnetic field lines can produce bright regions known as plages\revise{, which tend to spatially coincide with smaller-area photospheric faculae, introducing} further perturbations. The behavior of these chromospheric regions is complicated: not only do they exhibit more complex center-to-limb brightness variations, but they can be upwards of an order-of-magnitude more extended in area than spots \citep[e.g.,][]{Chapman2001, Meunier2010}. Indeed, posited planets have been shown to actually be stellar activity \revise{masquerading as Keplerian signals} (e.g., \citealt{Robertson2014} and potentially \citealt{Jeffers2014}). In one especially insidious case, stellar activity coupled with the observing window function led to the false detection of an Earth-mass planet in the $\alpha$ Cen B system \citep{Dumusque2012, Rajpaul2016}.  \par

Approaches to mitigating spurious velocities introduced by stellar activity have been multifaceted. In the observational realm, suspected planets with orbital periods near integer multiples of the stellar rotation period are generally disregarded since spot groups can persist for multiple stellar rotations and inject power into harmonics of the rotation period \citep{Robertson2014}. Recent astrophysical insights into radial velocity ``jitter" have also produced guidance for identifying and prioritizing stars with minimal intrinsic radial velocity variability \citep{Luhn2020}. To combat activity-induced noise in observed stars, astronomers have used different spectral lines, such as \Halpha\ \citep[e.g.,][]{Kurster2003, Hatzes2015} and the Ca H\&K doublet \citep[e.g.,][]{Wilson1968, Wright2004, Isaacson2010}, to simultaneously diagnose activity in stellar photospheres and chromospheres. However, these activity indicators are not universally exploitable: diagnostic lines will show variability among stars of differing spectral types, and even different tracers may differ on the same star. Often, different indicators must be calibrated to suit specific instrumental spectral ranges, especially in the case of M dwarfs \citep{Robertson2016}. \par

Recently, astronomers have begun applying advanced statistical models to RV data series in an attempt to disentangle activity from true Doppler shifts. Gaussian Process (GP) models for stellar variability have shown particular promise in separating out these two effects \citep[e.g.,][]{Haywood2014, Rajpaul2015,Jones2017}. \revise{Indeed, a large blind test conducted in \citet{Dumusque2017} suggested that GP-based analyses provided a promising way to model correlated signals due to stellar variability and could assist with extracting short-period planetary signals with low RV semi-amplitudes in the presence of stellar variability \citep{Dumusque2016}.} To aid in the \revise{further} development of these novel statistical methods, the community has developed algorithms to simulate the photometric and spectroscopic impacts of magnetically-active regions present on the stellar photosphere \citep[e.g.,][]{Herrero2016, Boisse2012, Dumusque2014}. The respective tools produced in these publications, \texttt{StarSim} and \texttt{SOAP 2.0}, have seen widespread use, since they are able to efficiently produce simulated data sets such as those used in the development of GP models \citep[e.g.,][]{Rajpaul2015, Jones2017, Gilbertson2020}. \par

Astronomers have also used our nearest star, the Sun, as a tool to understand the RV impact of individual magnetic features. For example, \citet{Haywood2016} use simultaneous observations of sunlight reflected off the asteroid 4/Vesta and full-disk solar observations from NASA's Solar Dynamics Observatory (SDO) to show that observational proxies for stellar plage coverage may enable better inference of activity-induced RV variations. Follow-up direct observations of the Sun with a dedicated solar telescope \revise{feed into the HARPS-N spectrograph \citep{Dumusque2015, Phillips2016, CollierCameron2019}} have reinforced this finding \citep{Milbourne2019}. \par

Although astronomers have made strides toward understanding activity from observational, theoretical, and statistical standpoints, novel approaches will be required to overcome another principal obstacle to centimeter-per-second precision RVs: granulation. Granules, individual magnetoconvective cells which blanket the surfaces of cool stars, are comprised of hot and bright upwelling plasma, surrounded by rings of cooler, denser downwelling plasma known as intergranular lanes. The collective motion of granules and larger convective cells known as ``supergranules" can introduce RV perturbations on the order of meters per second on timescales ranging from minutes to days \revise{\citep{Rieutord2010, Rincon2018}}. Unlike spots and plages resulting from activity, these convective phenomenon are not rotationally modulated, and so do not exhibit the same quasi-periodic time dependence. Therefore, distinct statistical methods and observational techniques may be required to distinguish spectral variability due to granulation from true Doppler shifts. \par 

Although an effective treatment for granular noise has not been devised, the physical processes by which granules perturb spectral lines is well-understood. In disk-integrated flux, typical absorption lines for solar-type stars exhibit a characteristic ``C"-shaped asymmetry that is observable in the line bisector curve \citep{Stathopoulou1993}. This characteristic shape results from the distinct motions within granules: more light is emitted by blueshifted upwelling plasma in granules compared to the redshifted downwelling plasma in lanes. An asymmetric absorption line results when the light from these distinct components is combined \citep{Gray2008}.

In disk-resolved observations, solar line bisectors vary in shape with limb position (see Figure~\ref{fig:one} and \S\ref{data}). At disk-center, typical bisectors exhibit the characteristic ``C"-shape seen in disk-integrated light. Toward the limb, however, bisectors become distorted, more resembling a ``\textbackslash"-shape. This transformation is an effect of viewing angle: at disk-center, the line of sight passes down vertically through the convective cell, whereas at the limb, the granules are viewed in profile, sampling different line-of-sight velocities. Due to these effects and their short lifetime (of order minutes), granules induce spectral perturbations that vary with time in both disk-resolved and disk-integrated light. \par 

Thus far, attempts to quantify the precise RV impact of granulation-induced line perturbations have been conducted via complex magnetohydrodynamic (MHD) simulations of small patches of the stellar surface, as in \citet{Cegla2018, Cegla2019a}. Using their MHD \citep[from][]{Vogler2005} and radiative transfer \citep[from][]{Socas-Navarro2015} simulations, these authors synthesized realistic disk-resolved Fe \textsc{I} 6302 \AA\ absorption line profiles. In \citet{Cegla2019a}, the authors then tiled a sphere with simulated patches to model a star and reconstruct a disk-integrated line profile. Using this approach, they were able to show that diagnostics of line bisector shape strongly correlated with the size of the measured spurious RV shift. However, these results are based on the synthesis of a single line. Real stellar spectra contain a myriad of lines with varying formation heights, magnetic sensitivity, etc. Given the computational time required to first conduct the magnetoconvective simulations and then run the radiative transfer, this approach is not \revise{currently} feasible for generating full synthetic spectra with dense time sampling. \revise{Moreover, the amplitude of granular variability naturally depends on the assumptions and physical simplifications made in solving the MHD and radiative transfer equations. Likewise, the presence, strength, and configuration of magnetic fields can greatly impact the inferred RV RMS. For example, \citet{Sulis2020} find an RV RMS of $\sim$0.507\ $\ms$, similar to the value observed by \citet{Palle1999}. Yet with the inclusion of strong magnetic fields, \citet{Cegla2019a} find a much smaller amplitude of only $\sim$10$\cms$.} \par  

\input{captions/fig1.tex}

Although these sorts of MHD simulations have yielded great physical insight, developing practical solutions to address convectively driven RV noise will require \revise{new, complementary} approaches. As we have seen from other astronomers' solar observations and data-driven photospheric models \citep[e.g.,][]{Dumusque2014, Haywood2016}, we can glean a great deal of practical knowledge from codes designed to produce observationally-informed synthetic data sets. \revise{Indeed, \citet{Sulis2017} indicate that simulated data sets could be used to train detection processes that are normally strongly affected by granular noise.} Although \revise{simulation tools such as} \texttt{StarSim} and \texttt{SOAP 2.0} have had great success modeling line perturbations from stellar active regions, neither includes a comprehensive treatment of time-variable perturbations from granulation. \revise{Other past works have used the observed and simulated properties of granules \citep{Meunier2015} and supergranules \citep{Meunier2019, Meunier2020a} to assess the impact of these convective motions on exoplanet detection prospects, though they only provide velocities and not spectra.} To fill in \revise{these gaps,} we have developed an observationally-informed simulation tool to create synthetic spectra with time-resolved, granulation-induced variability. We dub this tool \texttt{GRASS}, the GRanulation And Spectrum Simulator. \revise{We envision that \grass\ may be useful for training and validating various strategies for reducing the effects of stellar variability on precision Doppler measurements. In particular, mitigation strategies that make use of line-shape information could be tested on data sets generated by \grass. } \par 

In the following sections, we describe \grass\ in detail. In \S\ref{data}, we summarize the input observations used to create our model spectra, and briefly comment on the output of the simulation tool. In \S\ref{methods}, we discuss the processes by which \grass\ creates synthetic spectra from the input solar observations. In \S\ref{validation}, we validate the \grass\ modeling approach by comparison to disk-integrated reference observations. In \S\ref{results}, we present some first results obtained from \grass\ single-line simulations. In \S\ref{discussion}, we comment on some potential uses, discuss future avenues, and also highlight the limitations and caveats of \grass. \par 

\section{Data} \label{data}

Rather than using MHD and radiative transfer simulations to produce synthetic spectra \citep[as in previous studies; e.g.,][]{Cegla2018, Cegla2019a}, we instead use disk-resolved, high-resolution observations of the Sun to create our synthetic disk-integrated spectra. This modeling approach offers a number advantages. Principally, we know that our model captures the most important physics precisely because it is based on real data. Although the current version of \grass\ explicitly excludes phenomena such as spots, this is in principle a solvable problem with additional observations or modeling. Our input observations are described below, in \S \ref{subsec:input_data}. In \S \ref{subsec:output_data}, we briefly describe the output of \texttt{GRASS}, reserving a more comprehensive description of our methodology for \S \ref{methods}. \par

\subsection{Input Data Description} \label{subsec:input_data}
Since line-shape variations driven by granulation vary with time and location on the disk, synthesizing our model disk-integrated spectra requires temporally and spatially resolved input data. To account for these variations in line shape, we use a set of solar observations originally published in \citealt{Lohner-Bottcher2019} (hereafter \citetalias{Lohner-Bottcher2019}) as a template for line synthesis. We refer to these data as the ``input data." These input data consist of reduced spectra covering the Fe \textsc{I} 5434.5 \AA\ line, from which we measure bisectors and widths as a function of depth. Ideally, these data would be derived from high-resolution observations conducted simultaneously across the solar disk. To our knowledge, however, no such data set exists and obtaining such data would require a special-purpose instrument. Relevant spectroscopic parameters and quantities for the Fe \textsc{I} 5434.5 \AA\ line are summarized in Table~\ref{tab:feI}. \par

\subsubsection{Summary of Input Data Observations}

\input{tables/spectroscopic.tex}

The observations of this Fe line were carried out as described in \citealt{Lohner-Bottcher2018} (hereafter \citetalias{Lohner-Bottcher2018}). The observations\revise{, consisting of 20-minute sequences,} were performed using the Laser Absolute Reference Spectrograph \citep[LARS,][]{Doerr2015, Lohner-Bottcher2017} at the German Vacuum Tower Telescope (VTT). LARS employs VTT's high-resolution echelle spectrograph in combination with a laser frequency comb (LFC) to achieve superb spectral resolution $R \sim \lambda/\Delta\lambda \sim 7 \times10^{5}$ with an absolute wavelength calibration accuracy of $\sim$$0.02$ m\AA\ (about $1\ \ms$) in the wavelength region of interest. \par 

To sample line bisectors along different lines of sight, observations were carried out at eleven positions along each of four directions from disk center, oriented along and perpendicular to the sky-projected solar rotation axis (see Figure~\ref{fig:one}). The positions are radially parameterized as $\mu = \cos(\theta)$, where the heliocentric angle $\theta$ is the angle subtended by the line of sight and the solar surface vertical. The parameter $\mu$ is therefore 1 at disk-center and 0 at the limb edge. Individual observations \revise{had a $\sim$10\arcsec\ field of view set by the fiber-coupling unit and} were performed with a pointing accuracy of $\sim$1\arcsec. \par

To average over supergranular motions which would hinder the precise convective blueshift measurements performed in \citetalias{Lohner-Bottcher2018} and \citetalias{Lohner-Bottcher2019}, observations at each limb position were quickly scanned over ellipses with area corresponding to the typical angular area of supergranules. \revise{The sizes and orientations of the ellipses were varied with position on the disk in order to account for the effects of foreshortening. At disk center, a circular scan pattern was used, with a total diameter of $\sim$30\arcsec\ accounting for the size of the fiber-coupling unit. Slightly away from disk center (from around $\mu = 0.95$ to $0.7$), where the amplitudes of $p$-modes dominate over motions from supergranulation, the scanning ellipses had total axes dimensions of about 40\arcsec$\times$10\arcsec\ (i.e., including the width of the fiber-coupling unit). Toward the limb ($\mu \leq 0.5$), where horizontal supergranular motions dominate over $p$-modes, the total ellipse axes dimensions were about 40\arcsec$\times$30\arcsec.} \par 

Observations were carried out in \revise{sequences of} $\sim$20 minutes in length, with individual exposures conducted at a cadence of 1.5 seconds and then binned into ten-exposure (15-second) bins. To avoid \revise{the systematic effect of changes in convective blueshift} introduced by active regions, simultaneous $G$-band images from the LARS context camera and recent magnetograms from the Solar Dynamics Observatory (SDO) Helioseismic Magnetic Imager \citep[HMI,][]{Pesnell2012, Scherrer2012} were used to ensure that only quiet Sun regions were observed. \revise{In the case of the HMI magnetograms, the automatically identified active region patches, which are flagged by the data reduction pipeline as pixels with an absolute line-of-sight magnetic field strength greater than 100 G, were used to reject active regions \citep{Hoeksema2014}.} Subsequent data reduction, including absolute wavelength calibration and corrections for barycentric motion and solar rotation, were performed as described in \citet{Lohner-Bottcher2017} and \citetalias{Lohner-Bottcher2018}. For lines observed by \citetalias{Lohner-Bottcher2018} and \citetalias{Lohner-Bottcher2019}, the \revise{observed} wavelengths of solar lines were empirically measured \revise{with respect to the lab-frame wavelengths of Fe \textsc{I} lines to an accuracy of below $0.1$ m\AA, or about $2-4\ \ms$,} using the \revise{LARS LFC in combination with a} hollow cathode lamp. \par

\subsection{Output Data} \label{subsec:output_data}
As we describe in detail in \S \ref{methods}, we spatially interpolate and integrate the input data to create synthetic time series spectra like those for an unresolved point source. We refer to these spectra, and the associated cross-correlation functions and velocities, as the ``output data" of the simulation. \par

\section{Software Description and Methods} \label{methods}

\subsection{Program Language, Installation, and Performance}
The latest version of \grass\ and its associated documentation are available from GitHub\footnote{\url{https://github.com/palumbom/GRASS}}. \revise{The frozen version of \grass\ accompanying this work (v1.0) is archived on Zenodo \citep{GRASS}.} \grass\ is written entirely in \texttt{Julia}\footnote{\url{https://julialang.org}}, a high-level, performance-oriented language. To generate synthetic spectra, the user must specify some input parameters in a brief \texttt{Julia} script (see the associated documentation and tutorials on GitHub). In addition to the output spectra, users can choose to store derived data products such as cross-correlation function (CCF) profiles and/or apparent radial velocities measured from the spectra using one of several algorithms. \par 

\grass\ was designed to produce synthetic disk-integrated spectra more quickly than can be done through MHD and radiative transfer simulations. The compute-time and memory requirements of \grass\ are primarily dictated by three parameters: the grid size of the simulated disk ($N^2$, see \S\ref{subsec:res}), the number of time-steps simulated ($N_t$, with one step corresponding to one 15-second input observation), and the number of spectral resolution elements synthesized ($N_\lambda$). We find that the computation time scales quadratically with $N$, and linearly with both $N_t$ and $N_\lambda$. For this work, we simulate single-line spectra to validate \grass\ against reference disk-integrated observations. For these runs of \grass, we use \revise{$N^2 = 132^2$}, $N_t = 50$, and $N_\lambda = 193$ (corresponding to a spectrum 1.5 \AA\ wide with spectral resolution $R = 7\times10^5$ at $\lambda = 5434.5$ \AA). Depending on the exact hardware configuration, we find that we can generate 50 instances of output spectra in less than \revise{thirty seconds} on desktop hardware. \par

\subsection{Creating Synthetic Spectra} \label{subsec:create}
\grass\ is designed to create realistic time series of synthetic spectra with granulation-driven line-shape variation. The creation of these synthetic spectra from the input data is carried out in three broad steps, which we describe in the subsequent sections:

\begin{itemize}
    \item initial``pre-processing" of the input data, in which the reduced spectra are used to \revise{measure line bisectors and widths} as functions of depth within the line (\S \ref{subsubsec:preprocess}),
    \item simulation of a spatially-resolved stellar disk, in which the pre-processed data are used to re-create synthetic disk-resolved line profiles on a model stellar disk (\S\ref{sec:resolved_disk}),
    \item and spatial integration of the model stellar disk, in which the simulated disk-resolved spectra are co-added into a disk-integrated spectrum like those observed for point sources (\S\ref{sec:disk_int}).
\end{itemize}

The output data consisting of disk-integrated, time-series synthetic spectra can optionally be binned and used to compute cross-correlation functions and apparent radial velocities, as is typical for current RV surveys. We detail our procedure for these computations in \S\ref{subsubsec:obs} and \S\ref{subsec:veloc}. \par 

\subsubsection{Input Data Pre-processing}
\label{subsubsec:preprocess}

\input{captions/fig2.tex}

As described in \S\ref{subsec:input_data}, the input solar data are reduced spectra binned to a 15-second cadence. To uniquely specify line shape from the core to the wings, we directly measure line bisectors and widths as functions of depth from these reduced spectra. To do so, we isolate the section of the reduced spectrum containing only the observed line of interest (i.e., the Fe \textsc{I} 5434.5 \AA\ line), which we then interpolate onto an evenly-sampled intensity grid. The width at each depth is then calculated as the difference in wavelength between the red and blue sides of the line, and the bisector as the average. \par

In implementation, we perform a number of additional steps to produce smooth and accurate bisector and width measurements. We first average the flux across adjacent pixels in areas where the line profile is not smooth. To circumvent the impact of shallow line blends in the line wings, we model the far line wings above 93\% continuum level as Voigt functions. We use separate non-linear least squares fits of the red-ward and blue-ward wings (excluding pixels containing significant blends) to determine the best model parameters. Since the uncertainty in the bisector measurement increases where the slope in the line is small \citep{Gray2008}, we find that the top-most bisector measurements are quite inaccurate. To avoid the impact of these spurious measurements, we model the bisectors above 80\% continuum flux as the linear extrapolation of the bisector data measured between 70\% and 80\% continuum flux. \par 

For many limb positions, the full sequence of input data were not observed continuously. Often, several minutes of spectra were observed in one session, with the other session observed months later. In these cases, we adjust the mean velocity of the latter bisector series to match that of the first. For all results presented in this paper, we have performed additional simulations using only the longest set of contiguous data for each limb position. We find no significant difference in our results between these two sets of simulations. However, caution should nevertheless be taken when producing or interpreting series of spectra longer than $\sim$20 minutes. \par 

In some cases, we exclude highly spurious instances of the input data. We discard any epochs for which the bisector is more than $3\sigma$ removed from the mean bisector measurement at a given intensity and limb position. Due to apparent errors in the raw spectra reduction, we also entirely discard the data collected on 2017/10/17 for $\mu = 0.7$ on the northern solar axis. We instead only use the data for this position collected in 2016. Although this reduces the total time-baseline of available data, this section of the Sun only contributes $\sim$$4\%$ to the total disk-integrated flux in our simulations. \par 

Example pre-processed bisectors and width measurements are shown in Figure~\ref{fig:two}. Variability from pulsations and granulation are clearly visible in the bisectors. Whereas pulsations largely only change the horizontal position (i.e., velocity) of the bisector curves, granulation can change the curvature, shape, and position of individual bisectors. The line width measurements do exhibit a small degree of variability, though the level of variability is very small compared to that of the bisectors. \par 

The initial pre-processing described above is performed only once, and the resulting bisectors and width measurements are written to output files that are included in the \grass\ GitHub repository. Upon execution of \grass, the files containing these pre-processed data are read into memory and sorted into arrays that are assigned $\mu$ and axis \revise{(i.e., North, East, South, West with respect to the solar rotation axis)} identifiers. For each unique \revise{$(\mu{\rm,\ axis})$} pair, the corresponding array contains the full time-series of pre-processed bisectors and width measurements for that disk position. As detailed in the subsequent sections, these data are then used to generate synthetic absorption line profiles. \par

\subsubsection{Adjusting Input Line Depths} \label{sec:scale_chop}

Following from the detailed explanation of stellar line formation presented in \citet{Gray2008}, we \revise{can} use the information encoded in the deep Fe \textsc{I} 5434.5 \AA\ line to also model shallower lines. As explained in Chapter 17 of \citet{Gray2008} and shown in their Figure 17.14, the bisectors of shallower lines resemble the upper portions of the bisectors of deeper lines. This fact is driven by the nature of deep line formation in stellar atmospheres. Because the intensity in a given portion of a line is controlled by the source function at a corresponding physical depth in the stellar atmosphere, the wings of deep lines, which form deep in the photosphere, may be thought of as shallow lines themselves. Lines from species with identical lower energy-level abundance distributions should therefore share line profiles. Given this fact, we have written \grass\ to allow the user to generate synthetic lines of arbitrary continuum-normalized depths, assuming that the information encoded in the deep Fe \textsc{I} 5434.5 \AA\ line samples velocity structure along deep slant depths in the solar atmosphere. \par  

To model shallower lines, \grass\ must transform the pre-processed input data. In the case of the width measurements, \grass\ simply scales the data onto the new range of intensities for the shallower line. For the bisector measurements, \grass\ discards measurements for intensities below the depth of the line and interpolates the remaining points onto a same-length grid. \revise{Since the input line bisectors are extrapolated above 80\% of the continuum level, synthetic lines of depths less than 20\% are based solely on extrapolated data.} The consequences of this approach with respect to Doppler velocity measurements are analyze and discussed in \S\ref{sec:depth_var}. \par 

\subsubsection{Simulating a Stellar Disk} \label{sec:resolved_disk}
Following the initial pre-processing of the input data, \grass\ models a stellar disk by creating a uniform Cartesian $xy$ spatial grid of $N \times N$ pixels. The model stellar disk is described by a unit circle inscribed within the square grid. At each point on this model disk, \grass\ creates a synthetic, spatially-resolved spectrum consisting of any number of synthetic lines specified by the user. To create the synthetic line profiles, \grass\ first finds the nearest solar axis and value of $\mu$ to the point of interest on the grid. Then, it identifies the corresponding input solar observations and chooses a starting epoch from the input data. This epoch is chosen randomly for each spatial grid cell, which ensures that synthetic lines in separate spatial cells do not move in concert. \par 

On a given spatial grid cell, \grass\ synthesizes each absorption line by using the input data to assign the wavelength at each depth for the left- and right-hand sides of the synthetic line. The resulting absorption line profile, which is initially specified on a uniform grid of intensities, is then interpolated onto a uniform wavelength grid of the spectral resolution specified by the user. This process is repeated for any additional synthetic lines. When lines are sufficiently close (i.e., in the case of blends), the separate absorption profiles are multiplied to create a blended line profile.  \par 

After all lines have been synthesized within a given spatial cell, \grass\ applies a rotational Doppler shift based on the position of the cell on the disk. We use the differential rotation model from \citet{Snodgrass1990}, where the angular rotational velocity $\omega$ varies as a function of a solar latitude $\phi$:

\begin{align}
    \omega(\phi) &= A + B \sin^2\phi + C \sin^4 \phi \\
    A &= 14.713 ^{\circ}/{\rm day} \\
    B &= -2.396 ^{\circ}/{\rm day} \\
    C &= -1.787 ^{\circ}/{\rm day}
\end{align}

\noindent We use the best-fit values reported in \citet{Snodgrass1990} for the coefficients $A$, $B$, and $C$. The true rotational velocity $v_{\rm rot}$ at some latitude $\phi$ is then calculated as:

\begin{align}
    v_{\rm rot} &= v_0 / \omega(\phi) \\
    v_0 &= 0.000168710673\ {\rm R_*/day/}c
\end{align}

\noindent where $c$ is the speed of light, and the stellar radius $R_*$ is set to unity. This velocity is then projected onto the line of sight for the given sky-projected inclination of the rotation axis. The default inclination is given by the 3D unit vector $(0,1,0)$ corresponding to an equator-on star. We note that this calculation, as well as the spatial $xy$ grid, assume a spherical star (observed as a circle in projection). In the case of rapidly rotating stars, this assumption may not be valid. \par 

In addition to the effects of differential rotation, \grass\ must also account for limb darkening. To obtain $I_\mu$, the limb-darkened intensity at a given $\mu$, we use a simple wavelength-independent, quadratic limb darkening law from \citet{Kopal1950}:

\begin{equation}
\frac{I_\mu}{I_0} = 1 - u_1 (1-\mu) - u_2 (1-\mu)^2, 
\end{equation}

\noindent where the default values for the coefficients are $u_1 = 0.4$ and $u_2 = 0.26$, typical solar values for this wavelength region \citep[as verified via][]{Southworth2015}. We take $I_0$, the intensity at disk center, to be 1. For simulations of other wavelength regions or stars with different limb darkening behavior, the limb darkening coefficients may be easily changed by the user. \par 

\subsubsection{Integrating Over the Stellar Disk} \label{sec:disk_int}
As described in the previous sections, \grass\ essentially constructs the intensity as a function of time $t$, wavelength $\lambda$, and position on the disk $(x, y)$: $I(t, \lambda, x, y$).  To create a spatially-integrated spectrum for a point source $I(t, \lambda)$, we sum intensities over the spatial dimensions:

\begin{equation}
    I(t, \lambda) = C \sum_{i=1}^N\sum_{j=1}^N  I(t, \lambda, x_{i}, y_{j})
\end{equation}

\noindent where $C$ is a normalization constant such that the disk-integrated continuum level is normalized to unity when the individual disk-resolved spectra are summed over spatial dimensions. Positions outside the stellar disk contribute zero intensity. We compute $C = \pi / (2N^2)$, where $N$ is the length of the $N \times N$ spatial grid. \par

In implementation, we allocate only a $N_\lambda \times N_t$ array (where $N_\lambda$ is the number of spectral resolution elements and $N_t$ is the number of simulated observation epochs) and perform the spatial summation without storing the intensity at each location. This process circumvents the need to allocate a much larger $N\times N\times N_\lambda \times N_t$ array, significantly reducing memory overhead. \par 

We note that our method of disk simulation and integration destroys any coherency of $p$-modes that is naturally present in the input data. Since each spatial grid cell is initialized with a random starting epoch selected from the input data, the contribution of the $p$-modes to the RV RMS is effectively averaged over. Thus the RV RMS measurements presented in \S\ref{results} may appear lower than those that pulsations would be expected to produce. Because we are attempting to probe only the granulation-induced noise, this is actually desirable. \revise{However, we do note that an analysis of the power spectrum of synthetic RV time series produced by \grass\ (not shown) suggests that the peak residual pulsation amplitude persists at a level of $\lesssim$5$\cms$.} \par 

\subsubsection{Simulating Realistic Observations} \label{subsubsec:obs}
While \grass\ can synthesize spectra with arbitrary spectral resolution, the process described in the previous sections creates synthetic spectra at 
the same spectral resolution as the input data ($R \sim 7\times10^5$). Though these sorts of spectra may be useful for some purposes, they are not typical for current Doppler planet survey observations, which are conducted on instruments with $R \sim 50,000 - 150,000$. \par 

In order to degrade the spectral resolution to more typical values, we perform a convolution with a truncated Gaussian line spread function (LSF) with characteristic width $\sigma_{\rm LSF}$ computed from the desired lower spectral resolution. By default, the output spectra are degraded to $R \sim 1.17 \times 10^5$\revise{, comparable to the average spectral resolutions of the HARPS-N and NEID spectrographs}. As in real RV spectrographs, we oversample these spectra with about $\sim$6 pixels per spectral resolution element in the region of 5434.5 \AA. To simulate photon shot noise, we then add Gaussian white noise to each pixel in order to achieve a specified signal-to-noise ratio (SNR). In doing this, we assume that the Poisson noise is in the Gaussian regime. \par

By default, the synthetic spectra are generated at a 15-second cadence (the same as the input solar data described in \S\ref{subsec:input_data}). However, real observations of stellar spectra (i.e., for stars other than the Sun) use larger integration times and have practical constraints on the observing window function. For the analysis presented in \S\ref{subsec:obs_strat}, we compare synthetic observation patterns by defining parameters such as number of observations, exposure times, and any time between exposures (including readout and off-target time). We achieve this by appropriately weighting and binning the consecutive 15-second ``snapshots" produced by \grass. \revise{To save computation time and resources, spectra are not synthesized for simulated read-out or off-target times.} \par 

\input{captions/fig3.tex}

\subsection{Measuring Velocities from Synthetic Spectra} \label{subsec:veloc}
The spectral imprints of true planet-induced stellar velocities and granular velocities are distinct. Whereas Doppler reflex motions shift the stellar spectrum, granular motions perturb individual line shapes, creating skewness and kurtosis in their profiles. Although these perturbations should not necessarily be characterized or interpreted as wholesale velocities, these small effects are captured as an apparent source of stellar variability noise in cross correlation function (CCF) analyses. In order to explore the impact of granulation on these analyses, we have implemented a CCF-based algorithm \revise{from \citet{EchelleCCFs}} for measuring apparent RVs in the \grass\ package. \par  

The implementation of our CCF algorithm follows that of \citet{Pepe2002} \revise{and additionally accounts for the appropriate treatment of Doppler shifts discussed in \citet{Wright2019}.} We calculate the CCF by shifting a template ``mask" in velocity along each spectrum: 

\begin{align}
    {\rm CCF}(v_R) &= \int I(\lambda) \cdot M(\lambda_{v_R}) d\lambda.
\end{align}{}

\noindent where $I(\lambda)$ is the observed or simulated spectrum and $M(\lambda_{v_R})$ the mask at some velocity-shifted wavelength $\lambda_{v_R}$ given by the classical Doppler formula

\begin{equation}
    \lambda_{v_R} \approx \lambda_0 \left( 1 + \frac{v}{c} \right)
\end{equation}

\noindent where $\lambda_0$ is the rest-frame wavelength of the line. The mask $M(\lambda_{v_R})$ is the weighted sum of individual top-hat functions $M_i(\lambda_{v_R})$ centered on rest-frame line centers:

\begin{equation}
    M(\lambda_{v_R}) = \sum_i M_i(\lambda_{v_R}) \cdot w_i
\end{equation}{}

\noindent We take the weights $w_i$ to be the continuum-normalized depth of line $i$. For actual observations, the weights may also take into to account the instrument efficiency at the location of each line. \par 

Keeping with common practice, we shift the mask in velocity increments smaller than the spectral resolution of the synthetic spectrum to compute an over-sampled CCF. As in \citet{Pepe2002}, we compute a velocity from the CCF by fitting with a Gaussian model. We first estimate a full-width at half-maximum of the CCF profile, and then perform the Gaussian fit via non-linear least squares on the data around the CCF minimum. We find that fitting within $\pm$0.69$\sigma$ of the minimum reliably retrieves injected velocities for lines with widths typical of the Sun with errors well below 1 $\cms$. We report the mean of this Gaussian fit as the apparent Doppler velocity of the spectrum, which is expressed relative to the velocity zero point defined by the rest wavelengths of the lines in the CCF mask. \par 

\section{\grass\ Validation and Testing} \label{validation}

\subsection{Validating \grass\ against real solar observations} \label{subsec:compare_solar}

In order to validate and assess our line synthesis model, we compare our synthetic disk-integrated absorption line profiles and bisectors to those extracted from the IAG solar atlas \citep{Reiners2016}. To compare line profiles, we first use \grass\ to synthesize a disk-integrated absorption line profile for a model Fe \textsc{I} 5434.5 \AA\ line. We set the continuum-normalized depth to match that of the line in the IAG atlas. We then shift the synthetic and observed line profiles to the same rest-frame velocity. The two profiles are shown at left in Figure~\ref{fig:compare_solar} as the black and blue curves. The synthetic and IAG spectra are in general agreement, with deviations in flux between the two only as large as $\sim$5$\%$. We attribute these larger deviations to blends with shallow lines that were modeled out of the input data in the pre-processing stage (see \S\ref{subsubsec:preprocess}). These blends are more clearly visible in the residuals shown in the bottom panel of Figure~\ref{fig:compare_solar}. Any additional deviations are likely introduced by the presence of active regions in the IAG spectrum, which was observed from a different site during a period of heightened solar activity in 2014 \citep{Hathaway2015, Reiners2016}.  \par 

\revise{To more cleanly compare and evaluate our line synthesis model, we model line blends in the IAG spectrum by iteratively fitting Gaussian profiles to the largest deviations between the IAG and synthetic spectra. Although real stellar spectra will contain many blends between both stellar and telluric lines, we nonetheless choose to model a ``clean" line profile in this work for three primary reasons. Firstly, our input data were observed at lower SNR than the IAG spectrum, meaning that line blends in these spectra are not strongly detected. We find that smoothing over the line wings in the preprocessing stage (\S\ref{subsubsec:preprocess}) allows greater numerical accuracy in bisector and width calculations. Secondly, telluric line blends will naturally vary with observing time and location, so we should not expect those tellurics seen in our input data to match those observed by another instrument. Thirdly, we prefer the approach of using a single, ``clean" template line to create synthetic spectra. Given this template line, \grass\ can create simulated blended profiles of arbitrary overlap.}

The cleaned IAG spectrum is shown at left in Figure~\ref{fig:compare_solar} as the green dotted curve, with residuals plotted as green points in the bottom panel. \revise{We find that cleaning the IAG spectrum by} dividing out the best Gaussian fit for each line reduces the largest flux deviation \revise{between the spectra} to the $\sim$0.5$\%$ level. Indeed, in the top panel of Figure~\ref{fig:compare_solar}, the curves for the cleaned IAG and synthetic spectra essentially overlap. \par

To further validate our model, we compare CCF bisectors for the synthetic and IAG line profiles. In each case, the CCF is computed using a single top-hat mask entry at the rest wavelength of 5434.5232 \AA. We then measure a bisector from each CCF using the same method described in \S\ref{subsubsec:preprocess}. The bisectors are arbitrarily shifted in velocity to achieve maximum alignment to aid in visual comparison. These bisectors are shown at right in Figure~\ref{fig:compare_solar}. The synthetic (black) and IAG (blue) bisectors show remarkable agreement from $\sim$20\% to $\sim$60\% of the continuum-normalized flux. Above $\sim$60\%, the presence of the aforementioned line blends move the observed solar bisector artificially to the right. Comparing the synthetic (black) and cleaned IAG (green) curves, the improvement in agreement is notable. In all cases, we only compute bisectors up to 90\% flux, since the measurement uncertainty in the bisector computation inflates drastically above this level. \par 

\subsection{Effects of Spatial Resolution} \label{subsec:res}

\input{captions/fig4.tex}
\revise{To create synthetic spectra, \grass\ creates disk-resolved line profiles on a uniform spatial grid (as described in \S\ref{sec:resolved_disk}).} The \revise{ideal resolution of this spatial grid over the stellar surface would match that of the input observations. However, because the spatial resolution of the input observations was designed to approximate the typical angular area of supergranules and consequently varies with position on the disk (as described in \S\ref{subsec:input_data}), simulating an ideal grid is impractical.} Therefore, we choose \revise{a rectangular grid and set} $N$ \revise{(the number of grid cells along each Cartesian $xy$ direction in the sky plane) such that the area of the spatial grid cell matches the weighted-average area of the input data observations. Given the sizes of the elliptical oscillations described in \S\ref{subsec:input_data} and \citetalias{Lohner-Bottcher2018}, we compute $N^2\sim132^2$ as our optimized number of grid cells.} \par

\revise{Given the complexity of the spatial scanning patterns in the input data, we further validate our choice of spatial resolution by comparing the RV RMS of synthetic time series to direct measurements of the granulation RMS in the literature.} \revise{To perform this comparison,} we repeatedly generate time series of synthetic spectra consisting of a single absorption line at several spatial grid sizes. These noiseless spectra were generated at the full 15-second cadence of the input solar data and with the native spectral resolution of $R=7 \times 10^5$. The single synthetic line was centered on $5434.5$ \AA\ and was assigned a continuum-normalized depth of 0.8, comparable to the depth of the template line in the input data. We then measured velocities from these spectra using the CCF method described in \S\ref{subsec:veloc}. This process was performed many times to measure a robust mean and sample variance of the mean-adjusted RMS of the RV time series. \par

The results of this test are shown in Figure~\ref{fig:res}. We find that the RMS of the measured velocities falls off as a power law with a best-fit index of \revise{$\alpha \sim 1$ (dashed gray curve)}. Ideally, our \revise{chosen grid size should produce a synthetic RV RMS} similar to that observed in the real Sun. Unfortunately, as summarized by \citet{Cegla2019a}, this value is not precisely known and varies depending on both the line measured and time of observation. In Figure~\ref{fig:res}, we show the 2$\sigma$ confidence bands for \revise{two measurements of} the granulation RMS \revise{performed} by \citealt{Elsworth1994} ($31.9 \pm 9 \cms$, \revise{orange forward-hatched}) and \citealt{Palle1999} ($46.1 \pm 10 \cms$, green \revise{back-hatched}). These horizontal curves intersect our relation in the approximate region of $N = 2^7$ to $N=2^8$. \revise{Moreover, we find that our chosen grid size of $N^2 = 132^2$ (vertical dashed line) intersects the power law relation within the 2$\sigma$ confidence interval for the solar granulation RMS measured by \citet{Palle1999}, suggesting that our chosen $N$ produces physically valid levels of variability.} Given this \revise{validation}, we choose $N = 132$ (i.e., $N^2 = 132^2$) as the default \revise{length of our} spatial grid for the other simulations and analyses presented in this work. \revise{As a utility, we allow users to change the spatial resolution of simulations in \grass. However, the default value is set to $N^2 = 132^2$, and \grass\ reports a warning if another value is chosen.} \par

\section{Applications and First Results} \label{results}

\grass\ is a powerful tool for creating synthetic stellar spectra. By design, these spectra contain minimal magnetic activity-induced velocities and are free from planet-induced Doppler reflex motions. Ostensibly, the only sources of variability encoded in each epoch originate with granulation and pulsation. Therefore, by creating these spectra, we can assess in detail those perturbations created by granulation. In the following subsections, we analyze single-line spectra generated by \grass\ and offer preliminary observation advice for the mitigation of granulation noise. \par 

\subsection{Synthetic Line Depth and RV Variability} \label{sec:depth_var}
\input{captions/fig5.tex}

Given the asymmetric nature of stellar absorption lines \revise{and the trends in bisector shape seen with line depth \citep{Gray2008}}, it is conceivable that lines of different depths \revise{will} show varying levels of apparent Doppler variability. To test this possibility, we again generate many synthetic spectra consisting of single lines at different depths. We again add no noise to these spectra, and preserve the native spectral and temporal resolution of the input data. We plot the \revise{mean} RMS of the radial velocities measured from these spectra against line depth in Figure~\ref{fig:rms_vs_depth}. We find that the RV RMS scales with depth above a threshold of $20\%$ continuum flux. Lines shallower than $20\%$ continuum flux also show increased variability. However, we caution that these lines of lower depths are modeled solely from extrapolated input data \revise{(see Figure \ref{fig:two} and associated text)}, potentially compromising the reliability of these RV RMS measurements. \par 

\revise{Considering only lines deeper than 20\%, the total amplitude of the trend is only about several$\cms$. In principle, measuring radial velocities separately for shallow and deep lines may therefore be advantageous. However, even if one were able to achieve the desired signal to noise from lines of similar, shallow depths there would still be an RMS at the level of $\gtrsim$45$\cms$.} \par

\subsection{Effects of Inclination} \label{subsec:inc}

\input{captions/fig6.tex}

As an additional test of \grass, we examine the relationship between sky-plane stellar inclination and synthetic RV variability. As in the previous sections, we use \grass\ to generate many synthetic spectra at the native spectral and temporal resolutions of the input data. These spectra again consist of a single line at 5434.5 \AA\ with a continuum-normalized depth of 0.8, similar to the template line in the input data. As seen in Figure~\ref{fig:inclination}, the \revise{mean} RMS of the measured RV \revise{time series} shows no strong trend with sky-projected inclination. Additionally, we note no strong trend between the \revise{widths} of the RMS \revise{distributions} and the stellar inclination. This suggests that the effects of the rotationally-broadened CCF profile for equator-on stars does not significantly affect the amplitude of granulation-induced RV signals. \par 

\input{captions/fig7.tex}

\subsection{Informing Observing Strategy} \label{subsec:obs_strat}

In the previous sections, we generated evenly sampled, noiseless time series at the full 15-second cadence of the input solar observations. For real observations of stars other than the Sun, such data are impractical to obtain. Therefore, we use \grass\ to simulate more realistic observations to evaluate the effectiveness of various observation strategies at mitigating the impact of granulation. \par

To generate these spectra, we apply simulations of various observational limitations detailed in \S\ref{subsubsec:obs} for a variety of synthetic observation patterns. In each observation plan, we simulate a series of \revisetwo{300}-second observations (each corresponding to about \revisetwo{one full period} of the principal solar $p$-mode). Given practical limitations as well as the $\sim$20-minute baseline of our input data, we only simulate up to four \revisetwo{300}-second observations per night (i.e., \revisetwo{20} total minutes of exposure time) \revise{to avoid excessively cycling through the input data.} In addition to varying the number of observations, we also vary the time between observations. We use either a short wait time of 30 seconds (corresponding to typical readout \revise{times for current-generation optical RV spectrographs}), or a long wait of \revise{60 minutes}. \par 

As in the previous sections, we generate our single-line synthetic spectra at the native resolution of $R=7\times10^{5}$, but we now convolve the observed spectrum with a truncated Gaussian to degrade the spectral resolution to $R \sim 117,000$ (see \S\ref{subsubsec:obs}). This spectral resolution is comparable to those seen in the current generation of optical RV spectrographs. \par 

We additionally add simulated photon noise to achieve an SNR of \revise{$\sim$5000} per pixel. In realistic observations, the SNR per pixel is often closer to $\sim$150 \citep{Gupta2021}. Following \citet{Murphy2007}, the error in velocity for a single line at this noise level is of order meters per second. In real spectra, this is overcome by the use of many thousands of lines in the CCF computation. For this work, rather than generating much larger spectra, we instead only explicitly synthesize one line. Thus we choose to scale the SNR to the level such that the single-measurement velocity precision is \revise{$\lesssim$30 $\cms$}. To verify our choice of SNR, we repeated our analysis with several higher SNR values (not shown) and found no discernible differences in the distributions of measured velocities, suggesting that the impact of granulation noise dominates over photon noise in this high-SNR regime. \par

For each simulated observation plan, we measure apparent Doppler velocities by fitting to computed CCF profiles. We repeat this procedure \revisetwo{a few thousand times} in order to sample a distribution of measured Doppler velocities for each synthetic observation plan. The results of this experiment are plotted in Figure~\ref{fig:obs}. As expected, the distributions becoming increasingly narrow with increasing number of observations. \revisetwo{However, we note no significant difference in the widths of the short-wait and long-wait distributions for all observation plans. Indeed, simple two-sample Kolmogorov-Smirnov (KS) tests fail to reject the null hypothesis that the short- and long-wait distributions are drawn from the same parent distribution (see Table \ref{tab:obs} for a summary of the distribution statistics).} This result is not necessarily expected. Previous investigations into empirically-motivated observations strategies have suggested that returning to a star multiple times over the course of a night \revisetwo{(in a manner akin to our long-wait observation simulations)} leads \revisetwo{to a lower scatter in the measured velocities} \citep[e.g.,][]{Dumusque2011, Meunier2015, CollierCameron2019}\revisetwo{, whereas these distributions suggest no such result.} \par 

\revise{The short baseline of our input data is likely responsible for this apparent discrepancy. Because our simulations are based on 20-minute observation sequences, \grass\ is completely insensitive to longer-timescale phenomena, principally supergranulation. Additionally, to produce the long-wait simulated observations, \grass\ must cycle through the same input data some number of times, possibly leading to spurious correlations in the measured velocities.} Potential avenues for addressing the limited time-baseline are discussed in \S\ref{subsubsec:timescale}. \par 

\input{tables/observations.tex}

\section{Discussion and Future Prospects} \label{discussion}
This paper presents \grass, a computational tool for generating observationally-informed stellar spectra with granulation-driven perturbations. We include these perturbations by using spectra of the quiet Sun observed at various positions on the disk to synthesize line profiles. These perturbations, when interpreted as bulk Doppler velocities (like in CCF analyses), are a significant source of RV noise in planet-search surveys. \par

\grass\ is an important new tool in the suite of already-existing RV-focused star simulators. Whereas tools like \texttt{SOAP 2.0} \citep{Dumusque2014} are concerned with magnetic activity and the associated suppression of convective blueshift, \grass\ is solely concerned with granulation noise on much shorter timescales. Unlike previous simulations of magnetoconvectively-driven RV noise \citep[e.g.,][]{Cegla2019a}, \grass\ is not based on underlying MHD and radiative transfer simulations. Instead its use of solar line observations allow it to simulate spectra much more quickly. \par

\grass\ will benefit from additional development. In the following sections, we briefly discuss some of these opportunities, as well as some potential applications and limitations of \grass.  \par 

\subsection{Implications for Observing Strategy} 
Our initial analyses of \grass-generated synthetic spectra indicate a correlation between RV RMS and line depth, suggesting that deeper lines contain more information about the range of convective motions present in the solar atmosphere. Our observation simulations, as expected, showed that a greater number of exposures produces a narrower distribution of measured velocities (i.e., a more robust mean). Somewhat unexpectedly, these simulations suggested that \revisetwo{the wait-time between observations does not impact the observed scatter in measured velocities.} 

{multiple closely-spaced ($\sim$30-second off-target time) exposures produce lower RV scatter than taking multiple distantly-spaced (\revise{60}-minute off-target time) observations.} \revise{This result is in apparent disagreement with past studies, which have suggested observations spaced by hours are favorable} \revisetwo{for reducing the RMS of RVs binned across nights} \revise{\citep[e.g.,][]{Dumusque2011, Meunier2015, CollierCameron2019}.} \revise{We suggest that the short time baseline of our input data is responsible for this discrepancy. Because we only have 20-minute sequences of input data, \grass\ cycles repeatedly through the input data in order to create the long-wait simulations, potentially introducing spurious correlations.} \par 

As we mention in \S\ref{subsubsec:timescale}, longer baseline input observations could alleviate this limitation and also enable analyses of longer-timescale phenomena, such as supergranulation and meridional flows (see \citealt{Cegla2019b} for a review). \revise{Although \citet{Makarov2010} and \citet{Meunier2020b} find that meridional flows may have large impacts on exoplanet detectability, these motions are variable on very long timescales (i.e., from months to years). It is therefore unlikely that these additional unmodeled motions contribute significantly to the apparent inconsistency of this result with past findings, which consider observations spaced only by hours.} Future investigations should also examine the relationship and any correlation between measured RVs and traditionally-used line shape diagnostics (e.g., bisector span, bisector inverse slope, etc.). \par

\subsection{Limitations of the \grass\ Tool} \label{subsec:limits}
\subsubsection{Modeling Stellar Spectral Types}
\grass\ specifically uses \textit{solar} observations to inform granular variability in spectra. Thus the applicability of this tool for spectral types beyond G2V may be limited. Studies of line bisectors for dwarf stars with convective envelopes have shown that the typical bisector shape is extremely sensitive to spectral type \citep[Figure 17.15 of][]{Gray2008}. This is especially true for stars hotter than the Sun; as effective temperature increases, the velocity span of the bisectors increases, and the bisectors themselves reverse in curvature \citep[Figure 17.16 of][]{Gray2008}. This change in bisector velocity span and curvature indicates a change in granulation that may be caused by the thinning of the surface convective zone toward earlier spectral types \citep{Kippenhahn2012}. Toward cooler effective temperatures, typical bisector velocity spans and shapes change less rapidly. Therefore \grass\ may be conservatively used to model stars later than G2V, but this should only be done with extreme caution and \revise{very} strong caveat. In the future, \grass\ could use line bisectors produced by MHD simulations of other stellar types as input \citep{Dravins2021}. In this way, we believe \grass\ is highly complementary to the suite of Sun-as-a-star MHD simulations. \par

\subsubsection{Modeling Other Lines}
In this work, we explicitly validate the efficacy of \grass\ for only the Fe \textsc{I} 5434.5 \AA\ line. Although the bisectors of shallower lines generally resemble the upper portion of bisectors of deeper lines, there is still some variability between lines. Currently, caution must be taken when simulating synthetic lines in spectral regions far beyond the spectral neighborhood of 5434.5 \AA. Future iterations of \grass\ may include observations of other solar lines. Input data derived from observations of solar lines with very different excitation potentials \revise{and/or magnetic sensitivies are two} especially important \revise{additions} that should be made. Following \citet{Gray2008}, species of different excitation potentials may have very different lower energy-level abundance distributions and therefore different line profiles. Including some new input data could therefore enable us to model a wider swath of new lines. \par 

In addition to the above variability, the strength of limb darkening also has a wavelength dependence. At shorter wavelengths, limb darkening is stronger, increasing the importance of the bisector shape near disk center. We have also not included the effects of any potential limb brightening, since we do not model any lines for which this is a concern in this work. For future iterations of \grass\ that include plages and chromospheric lines, limb brightening would then become an important modeling consideration \citep{Meunier2010, Dumusque2014}. \par 

We also emphasize that our template line is a magnetically-insensitive line (i.e., it has Land{\'e} g-factor $g_{\rm eff} = 0.0$). Magnetically-sensitive lines, which can exhibit Zeeman splitting and Zeeman enhancement in the presence of a sufficiently strong magnetic field, are known to have their bisectors influenced by magnetic fields \citep[e.g.,][]{Cavallini1988, Brandt1990}. These effects are unmodeled by the current version of \grass. \par

\subsubsection{Input data timescale} \label{subsubsec:timescale}

The current version of \grass\ uses 20-minute baseline input observations. Sampled at a 15-second cadence, these observations are well-suited for studying the short timescales relevant for solar-type granulation, but fail to capture longer-timescale magnetoconvective phenomena. Future versions of \grass\ could take advantage of longer baseline input data to include the contribution of longer timescale phenomena, such as supergranulation. To the authors' knowledge, such long-scale data sets do not exist. If such observations are impractical for direct observations, MHD simulations could provide an alternative path to the requisite input data. Although MHD simulations are computationally expensive compared to \grass, these two methods could be valuable complements to each other. Given a single realization of the appropriate MHD-generated spectra, \grass\ could then produce many other realizations at a fraction of the computational cost. \par

\section{Concluding Remarks}
\grass\ is a new tool which uses the first empirical model of convective line-shape perturbations to generate stellar spectra with granular variability, as opposed to previous models which have relied upon more expensive magnetohydrodynamic and radiative transfer simulations. In this work, we have shown that \grass\  is a valuable and useful new tool for disentangling stellar RV variability from true Doppler shifts:

\begin{itemize}
    \item \grass\  accurately reproduces disk-integrated solar line profiles within $\sim$0.5\% and bisector profiles within a few meters per second (\S\ref{subsec:compare_solar} and Figure \ref{fig:compare_solar}). 
    \item \grass\ can probe the effects of line depth, stellar inclination, and other variables on apparent RV variability (\S\ref{sec:depth_var} and \S\ref{subsec:inc}).
    \item \grass\ in principle is useful for designing and evaluating observation strategies for minimizing noise from granulation (\S\ref{subsec:obs_strat}),
    \item \revise{but in practice, \grass\ is currently limited to timescales shorter than 20 minutes, set by the length of the input data which excludes signal from effects such as supergranulation and meridional flows.}
\end{itemize}

As a new entry in the suite of programs designed to model stellar spectra with Doppler variability, we hope that \grass\ will serve as a valuable tool in the road toward the detection and characterization of potentially Earth-like planets.

\acknowledgements
This research was supported by Heising-Simons Foundation Grant \#2019-1177. 
This work was supported by a grant from the Simons Foundation/SFARI (675601, E.B.F.). E.B.F. acknowledges the support of the Ambrose Monell Foundation and the Institute for Advanced Study. M.L.P. acknowledges the support of the Penn State Academic Computing Fellowship. The Center for Exoplanets and Habitable Worlds and the Penn State Extraterrestrial Intelligence Center are supported by the Pennsylvania State University and the Eberly College of Science. Computations for this research were performed on the Pennsylvania State University’s Institute for Computational and Data Sciences’ Roar supercomputer. This research has made use of the NASA Exoplanet Archive, which is operated by the California Institute of Technology, under contract with the National Aeronautics and Space Administration under the Exoplanet Exploration Program. This research has made use of NASA's Astrophysics Data System Bibliographic Services.  

\facility{Exoplanet Archive, VTT}
\software{\revise{\grass\ (v1.0; \citealt{GRASS})}, 
          EchelleCCFs (v0.1.11; \citealt{EchelleCCFs}), 
          \revise{Matplotlib (\citealt{Hunter2007})}}

\clearpage
\bibliography{main,misc}
\bibliographystyle{aasjournal}

\end{document}

%% file: captions/fig1.tex
\begin{figure*}[htb!]
\gridline{\fig{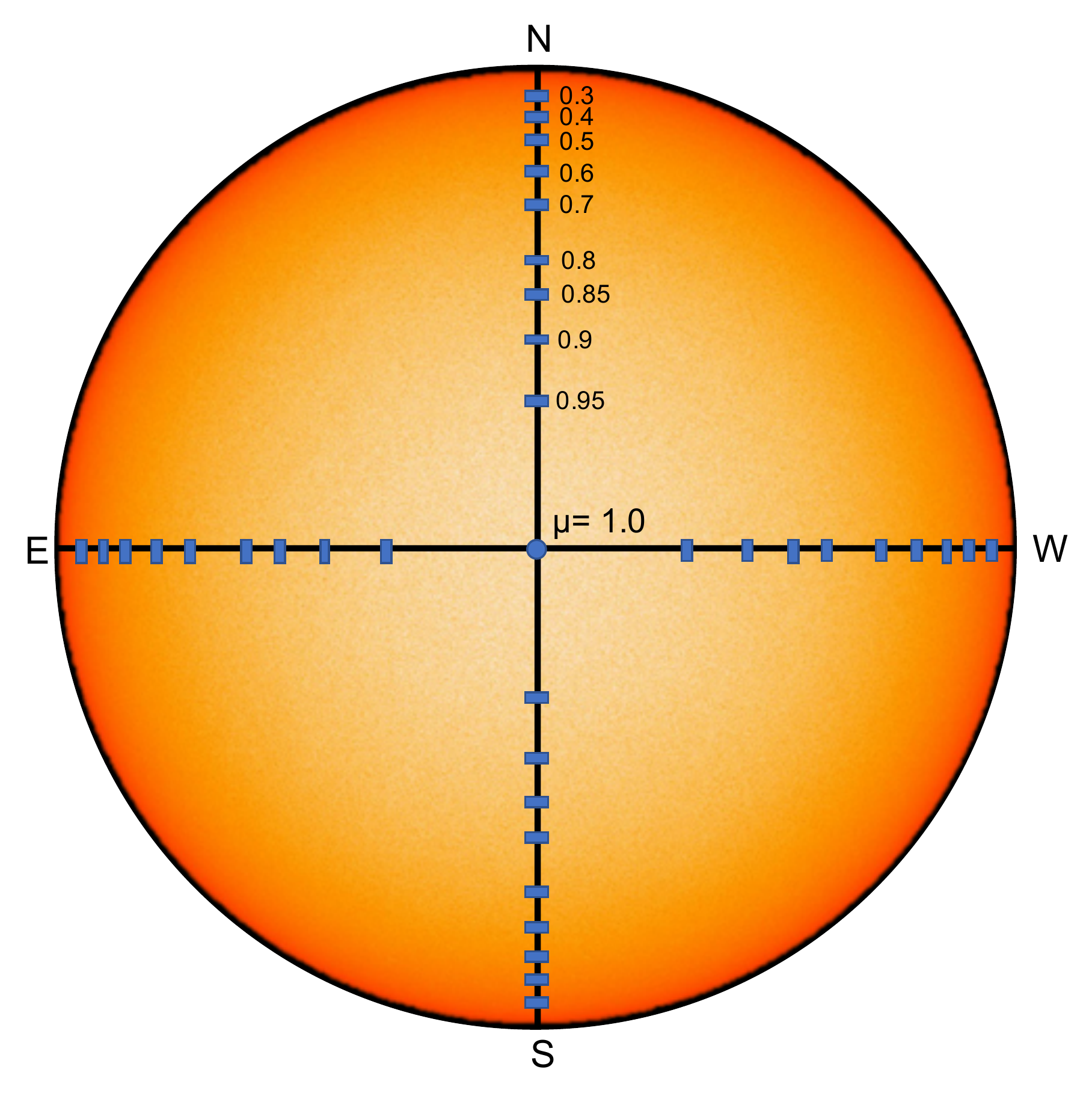}{0.4\textwidth}{}
          \fig{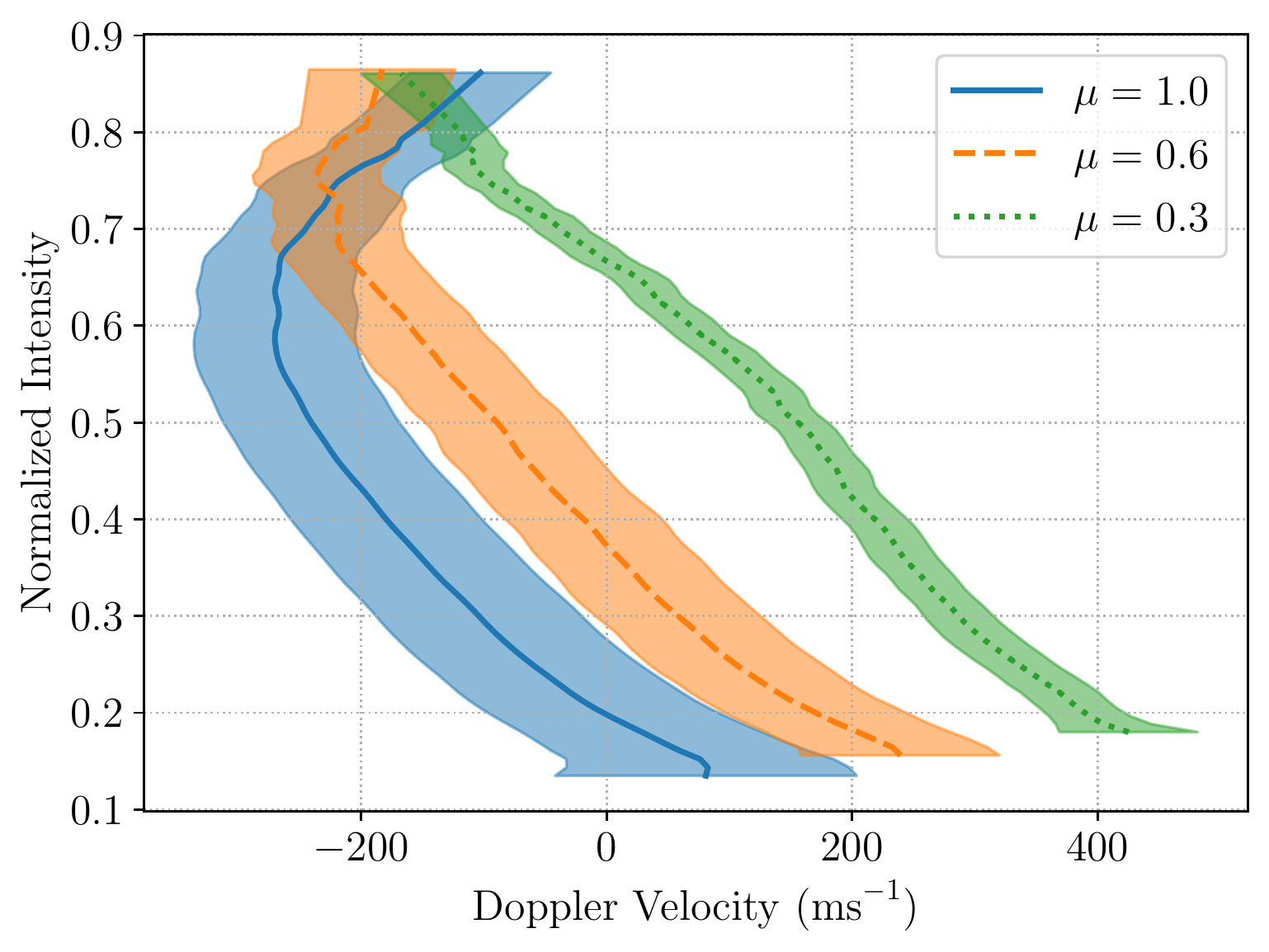}{0.55\textwidth}{}}
\caption{Temporal and spatial variability in input bisectors encode turbulent granulation velocities. \textbf{Left:} To capture the line-of-sight variation observed in solar bisectors, observations were carried out at many discrete disk positions along each of the four solar limbs plus disk center. Figure adapted from Figure 1 of \citetalias{Lohner-Bottcher2018}. \textbf{Right:} Line bisectors differ in shape with both time and apparent position on the solar disk. From disk center ($\mu = 1$) to the limb ($\mu = 0$), bisectors transition from a distorted ``C"-shape to something more resembling a ``\textbackslash"-shape. This behavior results from the different viewing angles of convective granules along different lines of sight \citep[Figure 17.13 of ][]{Gray2008}. The solid, \revise{dashed, and dotted} lines trace the time-averaged bisector for each limb position. The shaded regions correspond to the $\pm1\sigma$ temporal variability of the bisectors at each limb position. Error bars in velocity are not shown; with an absolute wavelength calibration accuracy on order $\sim$$0.02$ m\AA\ ($\sim$1 $\ms$), the observed temporal variations are much larger than the typical uncertainty. \label{fig:one}}
\end{figure*}

%% file: tables/spectroscopic.tex
\begin{deluxetable*}{ccccccccccc}[htb!]
\tablecaption{Parameters for spectroscopic lines in this work. Quantities denoted with a dagger ($^\dagger$) are from \citetalias{Lohner-Bottcher2019}. All other quantities are from the NIST Atomic Spectra Database \citep{NISTASD}. \label{tab:feI}}
\tablenum{1}

\tablehead{\colhead{Species} & \colhead{Air Wavelength (\AA)} & \colhead{Energy (eV)} & \colhead{Height (km)} & \colhead{$g_{\rm eff}$} & \colhead{$\log(gf)$} & \multicolumn{2}{c}{Spectroscopic Term} & & \multicolumn{2}{c}{Excitation Potential (eV)} \\
\cline{7-8}
\cline{10-11}
& & & & & & Lower & Upper & & Lower & Upper}

\startdata
Fe I & 5434.5232$^\dagger$ & 2.28078418 & 550$^\dagger$ & 0.00$^\dagger$ & -2.122 & 3d$^7$4s \ $^5$F$_1$ & 3d$^6$4s4p \  $^5$D$^{\rm o}_0$ & & 1.01105568 & 3.29183986 \\
\enddata
\end{deluxetable*}

%% file: captions/fig2.tex
\begin{figure*}
    \epsscale{1.1}
    \plottwo{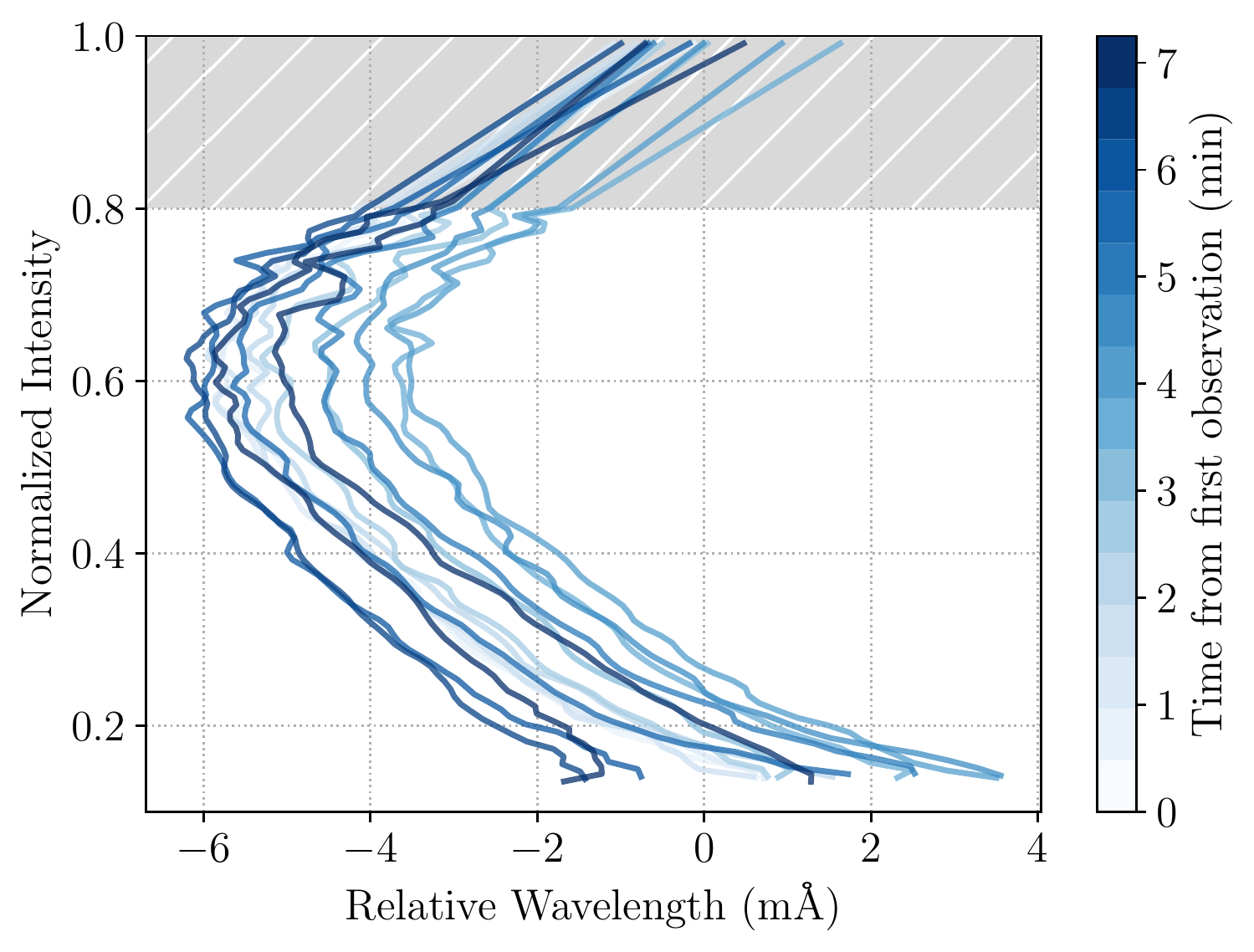}{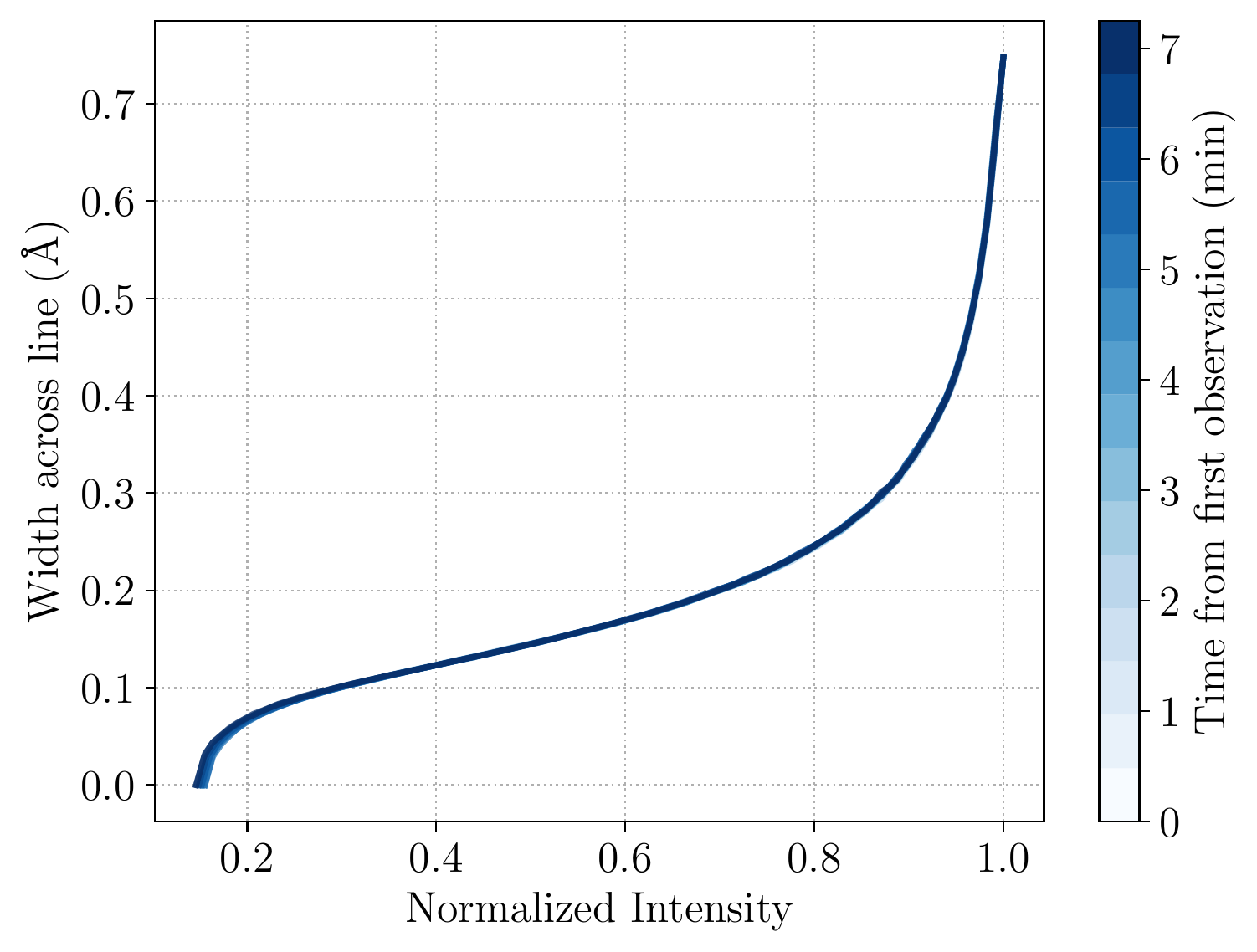}
    \caption{Example pre-processed input data showing granular variability at disk center ($\mu = 1$). Each curve shows the input data at one snapshot in time. The color gradient has been added to aid the eye in separating out these realizations. \textbf{Left:} Line bisectors show variation due to pulsations and granulation. Changes in bisector shape encode granulation, and changes in position result from the combined effects of pulsations and granulation. Above continuum-normalized flux of 0.8 \revise{(gray hatched region)}, we model the bisector as the extrapolation from lower flux values. \textbf{Right:} Line widths measured as a function of depth show \revise{only slight} variation with time. These width measurements display significantly less \revise{temporal} variability compared to the line bisectors. Together, line bisectors and width as a function of depth uniquely specify the shape of a line at a given instance in time.}
    \label{fig:two}
\end{figure*}{}

%% file: captions/fig3.tex
\begin{figure*}
    \epsscale{1.15}
    \plottwo{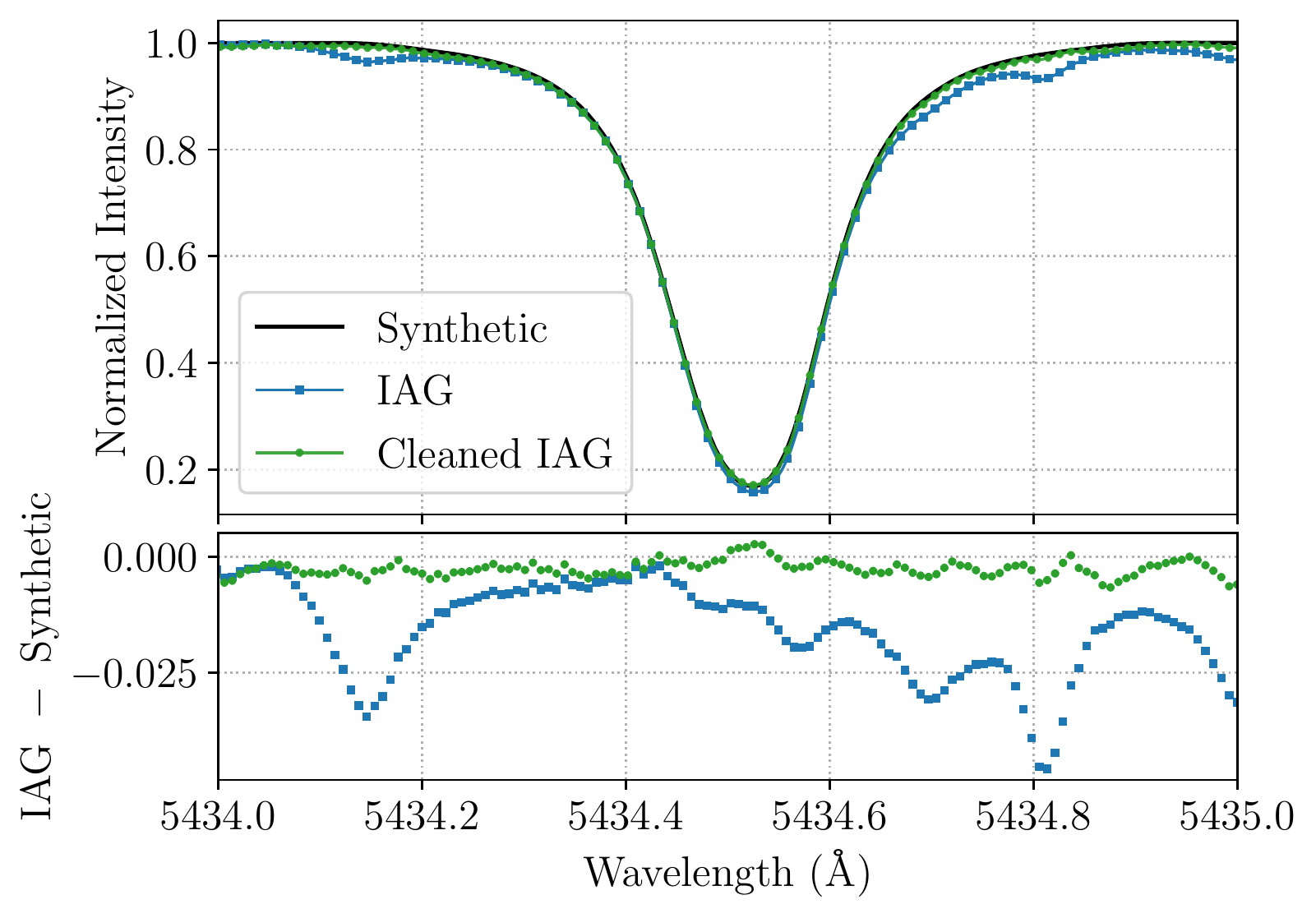}{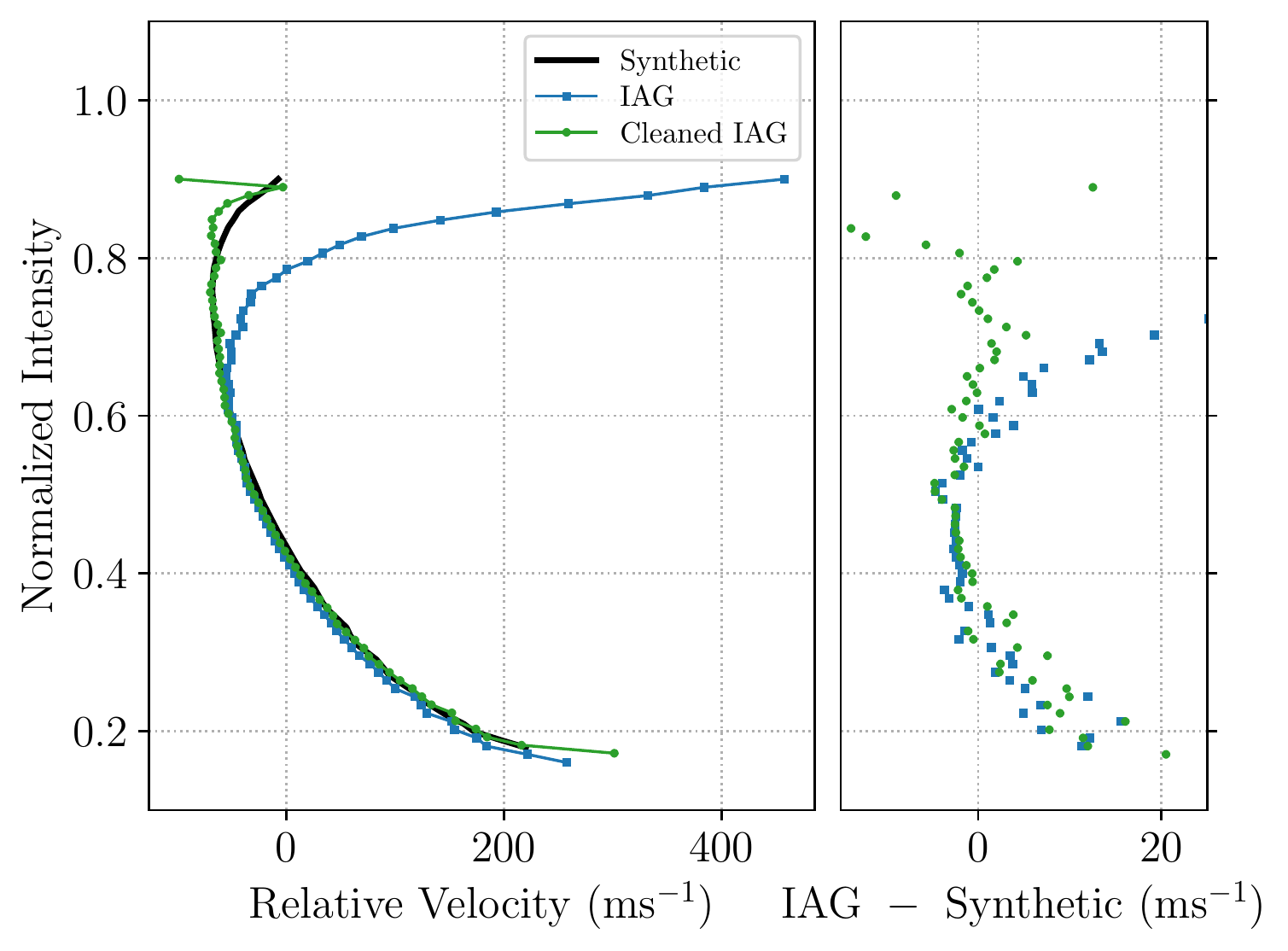}
    \caption{Validation of \grass\ line synthesis against the IAG solar atlas. \textbf{Left:} Observed (blue) and synthetic (black) Fe I 5434.5 \AA\ absorption lines in disk integrated light are overplotted. We additionally plot the IAG spectrum cleaned of shallow line blends in green, which nearly completely covers the black synthetic curve. Residuals between the IAG spectrum and the synthetic spectrum are shown in the bottom panel. Our synthetic spectrum deviates from the observed IAG spectrum at most by $\sim$5\% (blue squares), but only by $\sim$0.5\% from the cleaned IAG spectrum (green circles). \textbf{Right:} The observed IAG (blue) and synthetic (black) Fe I 5434.5 \AA\ bisectors, which have been arbitrarily shifted horizontally (i.e., in velocity) to achieve maximal alignment, generally agree in shape up to 60\% of the continuum level. The cleaned IAG (green) and synthetic bisectors are in much greater agreement. The residuals shown in the right panel further show that the blend cleaning greatly improves the agreement between model and observation, especially in the upper portion of the bisector.}
    \label{fig:compare_solar}
\end{figure*}{}

%% file: captions/fig4.tex
\begin{figure}
    \epsscale{1.2}
    \plotone{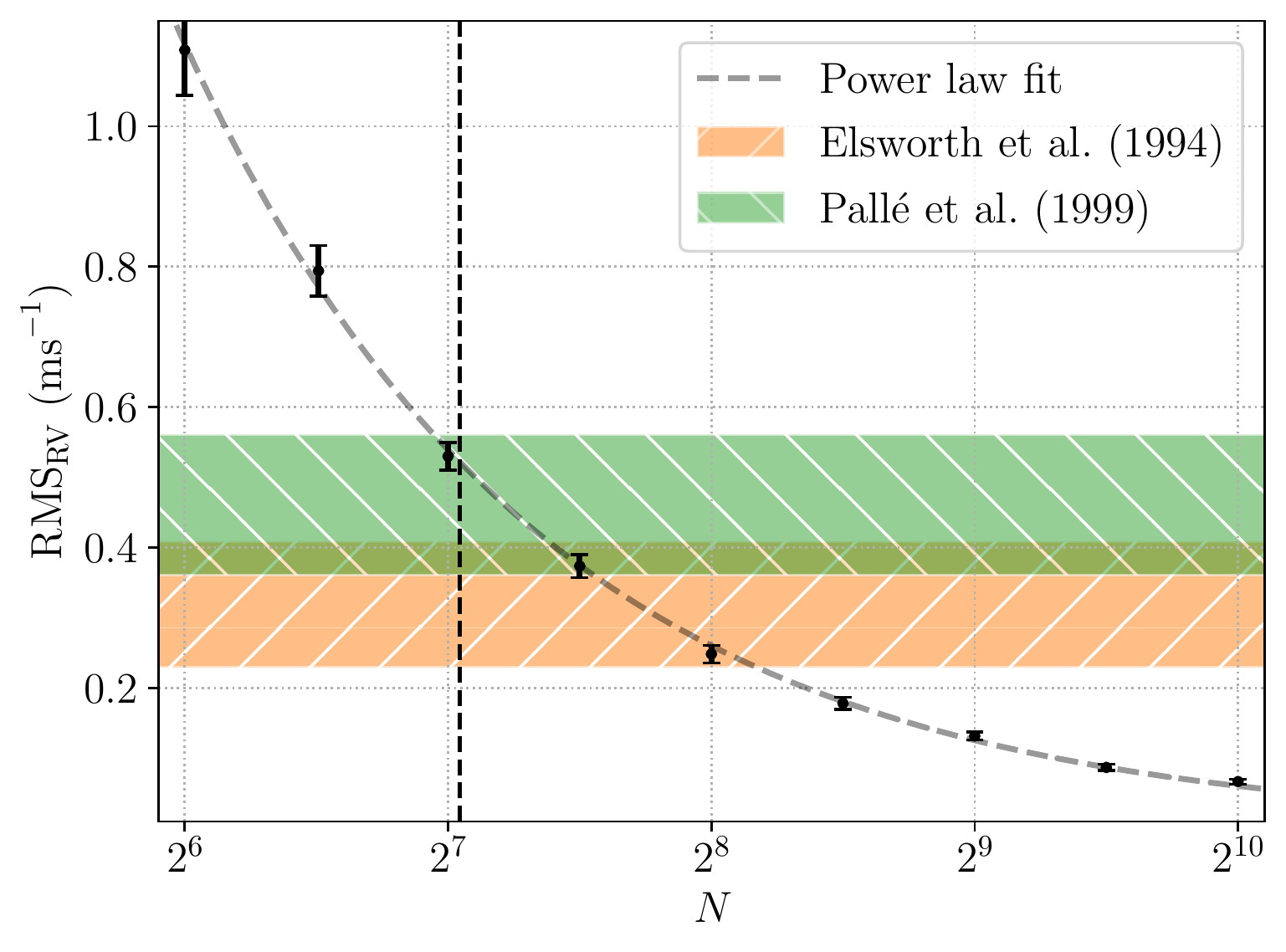}
    \caption{Effect of simulation spatial grid size on the generated synthetic spectra. Points indicate the mean RMS calculated from many RV time series, and error bars \revise{the standard error of the mean}. \revise{Two measurements of the RV variability from solar granulation are shown as the orange forward-hatched \citep{Elsworth1994} and green back-hatched \citep{Palle1999} regions. We find that our ideal grid size of $N^2 = 132^2$ (vertical dashed line), chosen to produce grid cells with angular area matching the weighted-average spatial resolution of the input observations, intersects the power law relation within the  \citet{Palle1999} region.} \revise{At spatial resolutions greater than $N\sim~400$, the angular size of the grid cells becomes much smaller than that of the smallest-footprint input observations and consequently loses physical validity.}}
    \label{fig:res}
\end{figure}{}

%% file: captions/fig5.tex
\begin{figure}
    \epsscale{1.2}
    \plotone{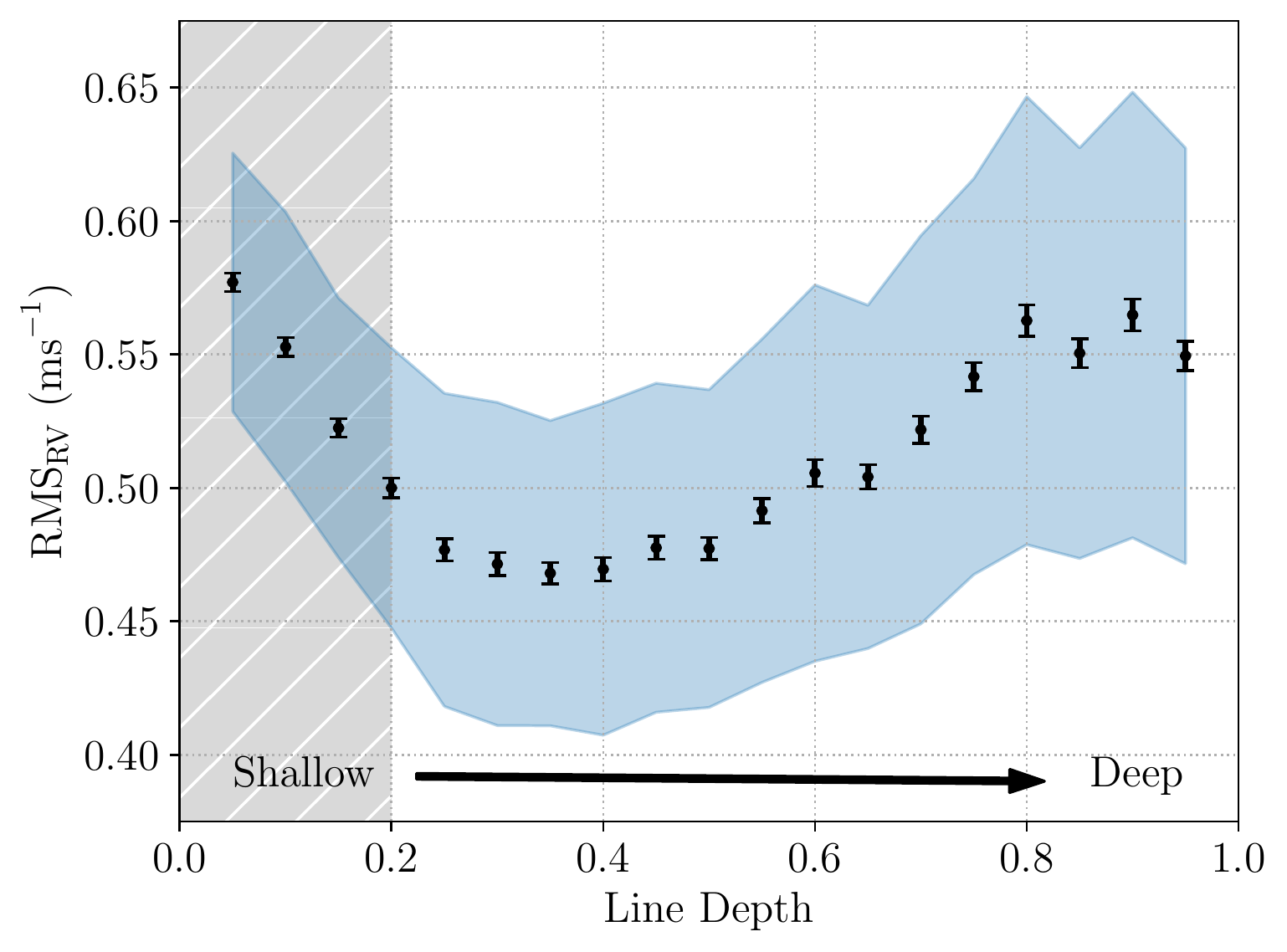}
    \caption{Doppler velocity variability weakly scales with synthetic line depth \revise{for lines deeper than 20\% of continuum level.} As in Figure \ref{fig:res}, points indicate the mean RMS calculated from many RV time series, and error bars the \revise{standard error of the mean}. \revise{The shaded blue region encloses 68\% of all RMS values across all runs.} From line depths of 0.5\% to 20\%, the RMS of an RV time series obtained from \grass\ output spectra appears to decrease. From 20\% to 95\% depth, the RMS then appears to steadily increase. However, the shallowest lines (gray \revise{hatched} region) are synthesized exclusively from extrapolated input data \revise{(see Figure \ref{fig:two} and associated text)}. The trend in this region may therefore be unreliable.} \label{fig:rms_vs_depth}
\end{figure}

%% file: captions/fig6.tex
\begin{figure}
    \epsscale{1.2}
    \plotone{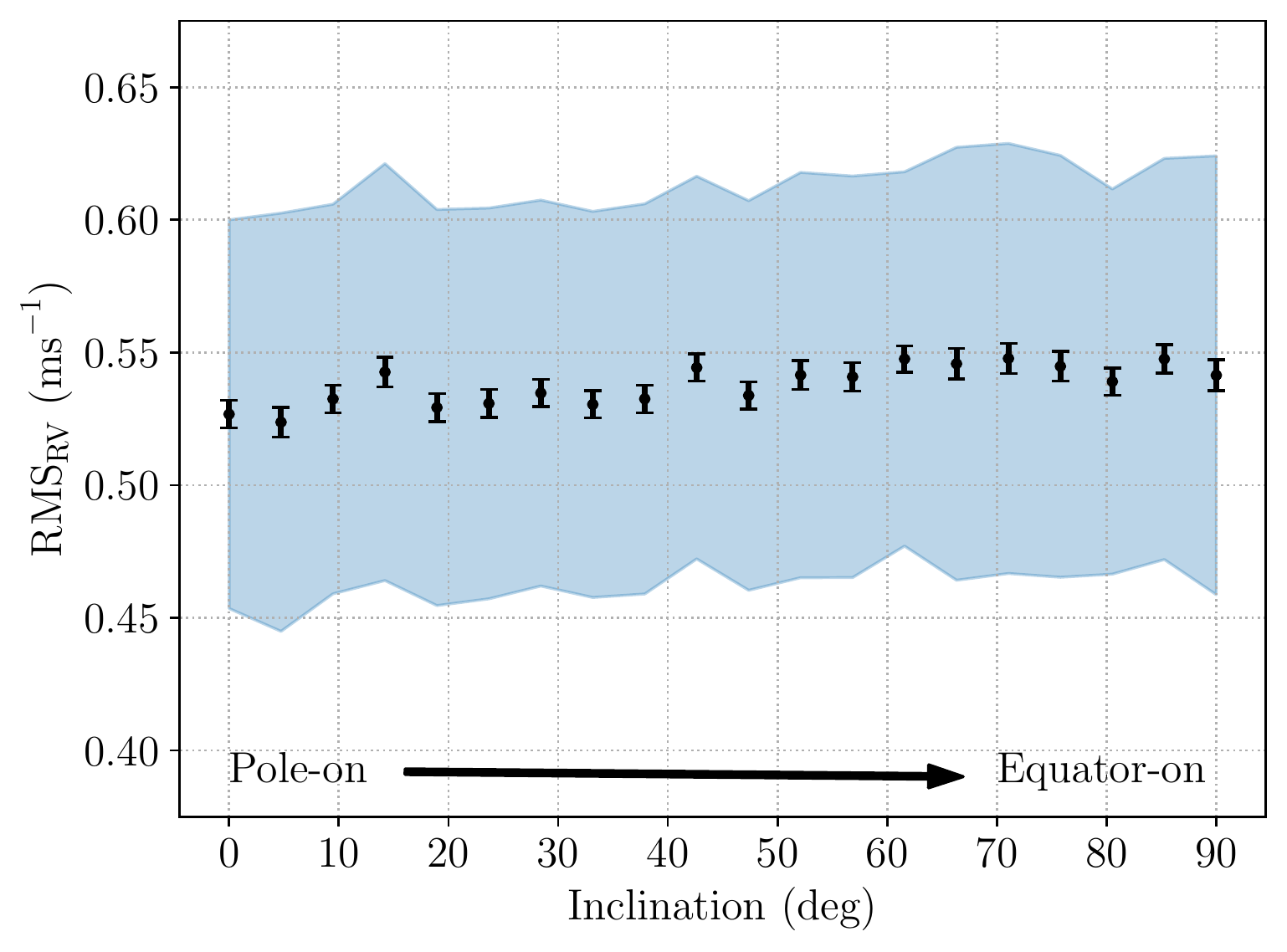}
    \caption{RV time-series variability and stellar inclination. As in Figure \ref{fig:res}, points indicate the mean RMS calculated from many RV time series, and error bars the standard \revise{error of the mean}. \revise{As in Figure \ref{fig:rms_vs_depth}, the shaded blue region encloses 68\% of all RMS values across all runs.} Equator-on stars do not appear to show any significant difference in \revise{granulation-driven} RV variability compared to pole-on stars. Additionally, \revise{neither the width} of the RV RMS \revise{distributions (blue shaded region)} \revise{nor the error on the mean (error bars)} appear to scale with inclination, suggesting that the impact of additional uncertainty introduced by a broader CCF profile for the Sun when viewed equator-on is \revise{small in comparison to the variability introduced by granulation signals.}}
    \label{fig:inclination}
\end{figure}{}

%% file: captions/fig7.tex
\begin{figure*}
    \epsscale{1}
    \plotone{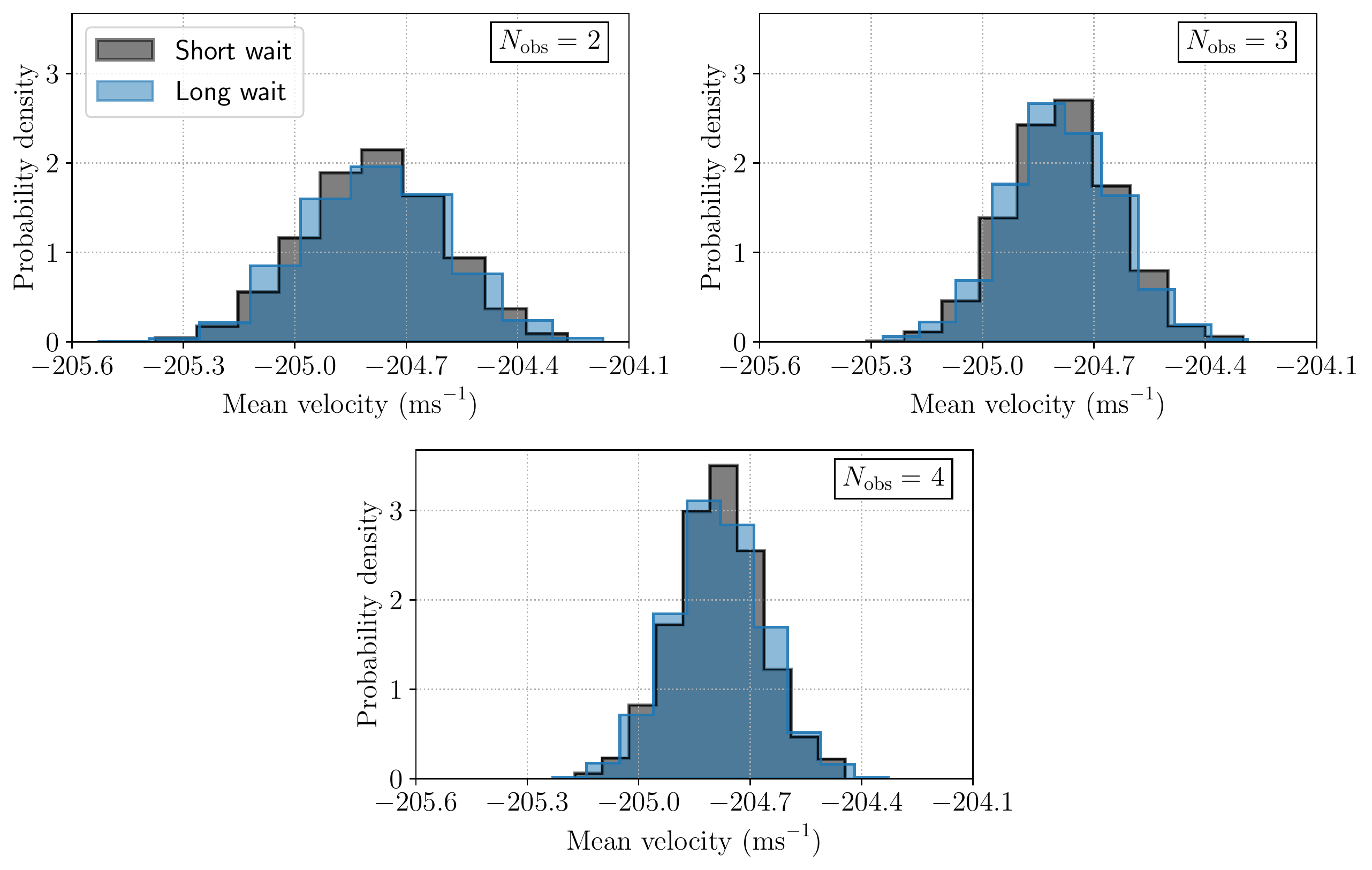}
    \caption{Distributions of \revise{mean} velocities measured from many realizations of synthetic observing programs. The number of exposures, each lasting \revisetwo{300} seconds, is varied between panels, as annotated in the upper right of each panel. \revise{For each given realization of a synthetic observation program, the velocities measured within the same night are averaged. Each distribution is built up from these averaged velocity measurements.} The black histograms correspond to simulated observing plans with short, 30-second lags between simulated exposures. The blue histograms correspond to long, \revise{60}-minute lags between simulated exposures. The \revise{relative} values of \revise{the} measured \revise{velocities} are themselves arbitrary, and correspond to the difference in rest-frame and CCF-template wavelengths. As expected, the distributions visibly narrow as the number of exposures is increased. \revisetwo{However, there is no significant difference in the width of the short-wait and long-wait distributions in all panels (see Table \ref{tab:obs}),} \revise{in apparent disagreement with past studies which have shown that long waits between observations produced lower scatter in measured RVs \citep[e.g.,][]{Dumusque2011, Meunier2015, CollierCameron2019}.} In the future, input data spanning longer than 20 minutes would allow a more robust comparison of observing strategies\revise{, since our short baseline input data do not preserve longer timescale variability (e.g., as in supergranulation)}.}
    \label{fig:obs}
\end{figure*}

%% file: tables/observations.tex
\begin{deluxetable}{ccccc}
\tablecaption{\revisetwo{Comparison of observing strategies with a short (30-second) or long (one-hour) wait between observations within a night (see Figure \ref{fig:obs}). Two-sample KS tests fail to reject the null hypothesis that the observed velocity distributions are drawn from the same parent distributions.} \label{tab:obs}} 
\tablenum{2}

\tablehead{\colhead{$N_{\rm obs}$} & \colhead{$p$-value} & \colhead{KS-statistic} & \multicolumn{2}{c}{RMS $(\ms)$} \\
\cline{4-5}
& & & Short Wait & Long Wait}

\startdata
2 & 0.578 & 0.779 & 0.184 & 0.191 \\
3 & 0.893 & 0.577 & 0.144 & 0.147  \\
4 & 0.441 & 0.866 & 0.117 & 0.125  \\
\enddata
\end{deluxetable}